\documentclass[prd]{revtex4}
\usepackage{amsfonts}
\usepackage{ulem}
\usepackage{amsmath}
\usepackage{amsmath,amssymb}
\usepackage{graphicx}
\usepackage{verbatim}
\usepackage{moreverb}
\usepackage[dvips]{color}
\usepackage{color}
\usepackage[vflt]{floatflt}
\usepackage{subfigure}
\usepackage{hyperref}
\usepackage{multirow}
\newcommand{\bea}{\begin{eqnarray}}
\newcommand{\eea}{\end{eqnarray}}

\newcommand{\bc}{\begin{Verbatim}%
[formatcom=\color{blue},frame=single,framerule=0.1mm,rulecolor=\color{black},xleftmargin=1.2cm,xrightmargin=1.0cm]}
\newcommand{\be}{\begin{equation}}						
\newcommand{\beql}[1]{\begin{equation}\label{eq#1}}					
\newcommand{\ee}{\end{equation}}
\newcommand{\beqn}{\begin{eqnarray}}
\newcommand{\eeqn}{\end{eqnarray}}	

\newcommand{\lept}[1]{\left\langle #1\right\rangle}

\newcommand{\qed}{\nobreak \ifvmode \relax \else\ifdim\lastskip<1.5em 
\hskip-\lastskip
\hskip1.5em plus0em minus0.5em \fi \nobreak\vrule height0.75em width0.
5em depth0.25em\fi}

\urldef{\PLANCKURL}\url{http://www.rssd.esa.int/index.php?project=Planck}
\begin{document}
\title{Voids as a Precision Probe of Dark Energy}
\author{Rahul~Biswas$^1$,
        Esfandiar~Alizadeh$^1$, and 
        Benjamin~D.~Wandelt$^{1,2,3}$}

\affiliation{$^1$ Department of Physics, 
University of Illinois at Urbana-Champaign,
1110 W. Green Street, Urbana, \\IL 61801, USA}
\affiliation{$^2$  
Department of Astronomy,
University of Illinois at Urbana-Champaign,
1002 W.Green Street, 
Urbana,\\ IL 61801, USA}
\affiliation{$^3$  
Institut d'Astrophysique de Paris, 98 bis bd Arago, France, CNRS/Universite Pierre et Marie Curie}
\begin{abstract}
A signature of the dark energy equation of state may be observed in
the shape of voids. We estimate the 
constraints on cosmological parameters that would be determined from 
the ellipticity distribution of voids from future spectroscopic surveys
already planned for the study of large scale structure. 

The constraints stem from the sensitivity of the distribution of 
ellipticity to the cosmological parameters through the variance of 
fluctuations of the density field smoothed at some length scale. 
This length scale can be chosen to be of the order of the comoving
radii of voids  at very early times when the fluctuations are 
Gaussian distributed. 
We use Fisher estimates to show that the constraints from void 
ellipticities are promising.
Combining these constraints with other traditional methods results 
in the improvement of the Dark Energy Task Force Figure of Merit on the dark energy 
parameters by an order of hundred for future experiments.
The estimates of these future constraints depend on a number of 
systematic issues which require further study using simulations. We 
outline these issues and study the impact of certain observational and 
theoretical systematics on the forecasted constraints on 
dark energy parameters.
\end{abstract}
\maketitle
\section{Introduction}
A number of 
observations have established that the expansion of the universe
is accelerating at late times
~\citep
{1999ApJ...517..565P,
1998AJ....116.1009R,
1998ApJ...509...74G,
2003ApJ...598..102K,
2003ApJ...594....1T,
2004ApJ...607..665R,
2006A&A...447...31A,
2007ApJ...666..694W,
2009arXiv0901.4804H}.
The cause of acceleration is usually attributed to an otherwise unobserved
component called dark energy, but models of dark energy are generically 
plagued by fine-tuning issues 
~\citep
{1989RvMP...61....1W,
1992ARA&A..30..499C,
2000astro.ph..5265W,
2001LRR.....4....1C,
2004mmu..symp..235C}.
One can also interpret these observations 
as a consequence of the gravitational dynamics being different from the 
evolution of a standard  FRW universe under general relativity. 
Such differences could  arise 
due to the symmetries of the FRW universe being broken in the real universe,
and the assumptions of smallness of the perturbations being invalid
~\citep
{2000PhRvD..62d3525B,
2005PhRvD..71b3524K,
2005PhLA..347...38E,
2006NJPh....8..322K}, 
or because General Relativity is not a correct description of gravity
~\citep
{2000PhLB..485..208D,
2005PhRvD..71f3513C,
2008PhRvD..78f3503J}.
With such fundamental questions at stake, a prime objective of 
physical cosmology is to understand the source and nature of this 
acceleration. 
All available current data 
~\citep
{2001ApJ...553...47F,
2005MNRAS.362..505C,
2006PhRvD..74l3507T,
2007MNRAS.381.1053P,
2008arXiv0803.0547K,
Dunkley:2008ie,
2008AJ....135..512O,
2008ApJ...686..749K,
2009arXiv0901.4804H}
is consistent with an FRW universe having 
dark energy in the form of a cosmological constant, yet various 
models of different classes are still allowed by the data. 
Therefore an important objective of
current and future observational efforts is to study the acceleration of
the universe in different ways and detect departures in the behavior from  
that expected in a standard $\Lambda$CDM  model. 

In order to compute parameter constraints from observational data, 
one usually parametrizes our ignorance about
dark energy with a time dependent equation of state (EoS) of 
dark energy as a specific function of redshift and theoretically computes 
the observational signatures. A very widely used choice, following the
recommendations of the Dark Energy Task Force
~\citep{2006astro.ph..9591A},
 is the CPL parametrization of the Equation of state 
~\citep{2001IJMPD..10..213C,2003PhRvL..90i1301L}.
This results in joint constraints on different parameters of the 
cosmological model, including the parameters of the EoS of dark energy.
It is important to use different sets of observational data. 
Different kinds of data sets probe different physical imprints of 
dark energy leading to distinct shapes of constraints on parameters. 
Consequently, the simultaneous use of many `complementary' probes 
leads to the tightest constraints on cosmological parameters
~\citep{1998astro.ph..5117T,1998astro.ph..4168T,1999ApJ...518....2E,2003PhRvD..67h3505F}. 

Moreover, as indicated above, we can hardly be certain that the specific 
parametrization of the EoS chosen, or even the choice of the physical 
model causing the  acceleration is correct. 
In that light, probing the observable effects of dark energy in terms of 
different physical aspects is even more important. 
A tension between constraints computed from different subsets of 
available data may be indicative of an incorrect parametrization 
~\citep{2006astro.ph.11178C}, or even an untenable choice of a physical 
model. 
Traditionally, the main observations used to constrain cosmological 
parameters have pertained to the apparent magnitudes of Type IA 
supernovae, the power spectrum of anisotropies of the Cosmic Microwave 
Background (CMB), and the power spectrum of inhomogeneities in the 
matter distribution (matter power spectrum). 
The constraints from the supernovae relate  to effects on the 
geometry of the universe due to dark energy through 
the changes in the background expansion. 
The CMB and matter power spectrum constraints stem mostly 
from a measurement of the geometry through the 
angular location of peaks of the anisotropy power spectrum and the peak 
positions of the Baryon Acoustic Oscillations (BAO), but also its
effects on the growth of perturbations through the magnitude of the 
power spectrum. Current status of the parameter constraints on the basis of
recent CMB, LSS, SNE observations can be found in~
\citep{2008PhRvD..77l3525W,2008PhRvD..78h3524X,2009arXiv0903.2532B}.
Further, the use of observations of clusters of galaxies and weak lensing
can be used to measure the growth of perturbations. It is therefore 
important to use probes of different aspects of cosmic evolution for 
constraining the cosmological parameters and models. 
From the viewpoint of both these perspectives, 
new probes for studying dark energy parameters are invaluable. 
 
In the above mentioned probes of the growth of cosmic structures, one 
studies the dependence of the dynamical growth of fluctuations on the 
cosmological parameters through the dependence of the 
growth of the amplitude (ie. size) of the fluctuations on the cosmology.
However, in standard cosmology, while the fluctuations are stochastically 
isotropic, the individual fluctuations are not isotropic. Thus, a measure
of the anisotropy and the time evolution of such measures can depend on 
cosmology in a distinct way. Consequently, this may be used to further 
constrain cosmological parameters. 
One expects that the signatures of anisotropic measures in observations 
would be related to the shapes of observed structures. 
Studying the evolution of shapes of high density regions 
(observable as galaxies or galaxy clusters at late times) and comparing 
with theory (eg.~\citep{2006ApJ...647....8H}) is difficult because this 
requires high resolution numerical simulations capturing the non-linear evolution of these systems.
This difficulty can be avoided to a large extent by studying voids using 
semi-analytic methods.
Therefore, the shapes 
of voids can be used to probe cosmology through the evolution of the 
anisotropy of fluctuations during cosmic growth.

Park and Lee~\citep{2007PhRvL..98h1301P,2007arXiv0704.0881L} 
identified the probability 
distribution of 
a quantity which they called ellipticity
\footnote{We note that this is not the conventional 
definition of ellipticity. Nevertheless, this is a convenient measure of 
the departure from spherical symmetry. Following 
~\citet{2007PhRvL..98h1301P}, we shall refer to it as the 
ellipticity in the rest of the paper}
related to the 
eigenvalues of the tidal tensor. 
They showed that 
the distribution was 
sensitive to the dark energy equation of state. Besides, they stated 
that the ellipticity could be derived from a catalog of galaxies, 
identifying
voids of different sizes and measuring their shapes, and the distribution 
was verified using results from N-body simulations. This ellipticity is
an example of a measure of anisotropy of individual fluctuations. 
The comparison of the probability distribution can provide
complementary constraints on dark energy parameters if its cosmology
dependence is different from other probes.
We will not require new probes to study constraints from voids,
rather one can study them using probes designed to study large scale 
structure in conventional ways, thereby allowing for better leveraging of
data.
Voids may be detected by the use of different void identification 
algorithms
~\citep{1997ApJ...491..421E,2001astro.ph.10449H,2002ApJ...566..641H,2008MNRAS.386.2101N,2008MNRAS.387..933C}, which find voids using different
characteristics, and may be considered to be different definitions of 
voids. Properties of voids have been explored
in 2dF~\citep{2004ApJ...607..751H} in SDSS~\citep{2005ApJ...621..643G,2007AstL...33..499T}. 
The shapes and sizes of voids in the SDSS DR5 have been explored in  
~\citet{2009arXiv0904.4721F}.

The main objective of this paper is twofold: (a) we want to quantify the 
potential of using void ellipticities to probe the nature of dark energy 
in terms of constraints on dark energy parameters,
(b) and to clarify the model assumptions that are important for this procedure, which should be verified, or modified according to results from simulations. 
This paper is organized as follows: In Sec. II we review the idea that
the shapes of voids can be quantified in terms of asymmetry parameters 
that can be related to the tidal tensor. We discuss the initial 
distribution of eigenvalues of the tidal tensor, and their evolution to
 study the evolution of the asymmetry parameters of voids and their
dependence on the underlying cosmology. 
There are different theoretical choices of models to approximate the non-linear evolution of the initial potential field to observable void ellipticities. We discuss two different choices in the appendix and show that our 
results are insensitive to these choices.
In Sec. III, we discuss the 
parameters from the surveys considered and our method of estimating the
number of voids identified from these surveys. In Sec. IV, we write 
down a likelihood and explicit formulae for the Fisher matrix and use them
to forecast constraints from these surveys. We also study how the 
constraints are degraded by systematic issues. We summarize the paper and 
discuss our outlook in Sec. V. 
\section{Theory}
In this section, we outline the basic idea of using asymmetry parameters 
describing the shapes of voids in estimating cosmological parameters. 
The anisotropy of fluctuations may be captured by
studying the eigenvectors and eigenvalues of the tidal tensor, which may be
 visualized as an ellipsoid with its principal axes along the eigenvectors
 of the tidal tensor, and sizes of the principal axes equal to the 
eigenvalues of the tidal tensor. At early times, 
the distribution of these eigenvalues at any point in space is known, and
their evolution can be studied by semi-analytic methods. Therefore, the 
distribution of these quantities may be computed theoretically and it is 
desirable to find observational signatures of this distribution.
Voids form around the minima in the density field of matter. The void 
geometry may be approximated by an ellipsoidal shape, which we shall 
refer to as the void ellipsoid. The central idea of 
~\citet{2007PhRvL..98h1301P} is that
the shape of the void ellipsoid as quantified by relative sizes of its
principal axes is set by the geometry (functions of the eigenvalues) 
of the tidal ellipsoid and these should be strongly correlated. 
This implies that the ellipticity measured from the geometry of voids can 
be used as an observable for specific functions of the eigenvalues of the tidal 
tensor. Observations of void shapes at different 
redshifts can then be used to trace the evolution of the stochastic 
distribution of these eigenvalues of the tidal ellipsoid at different 
redshifts. This contains dynamical information that may be used to 
constrain cosmological parameters. 

We briefly describe measures of ellipticity of the void ellipsoid and 
their connection to the eigenvalues of the tidal ellipsoid in 
subsection ~\ref{ansatz}: this specifies the functions of the
tidal eigenvalues that are constrained by the void shapes. 
We then describe the distribution of eigenvalues of the initial tidal 
tensor appropriate to an observed void in subsection 
~\ref{distribution_eigenvalues}. 
Then, in 
appendices ~\ref{appendix:gesf} and ~\ref{appendix:EllipsoidalCollapse}, we study the time evolution of 
the initial eigenvalues using two different approximations, and find them 
to be consistent. 
\subsection{Relating the Asymmetry Parameters to the tidal tensor}
\label{ansatz}
%
%
%
%
To describe the dynamics, we choose the comoving coordinates of 
particles (or galaxies)
as the Eulerian coordinates $\vec{x}$, while the Lagrangian coordinates
are taken to be $\vec{q}$, which are approximately the `initial' Eulerian 
coordinates at some chosen large redshift. The two coordinates are
always related 
through the displacement field $\vec{\Psi}(\vec{q},\tau)$.
\begin{equation}
	\vec{x} =\vec{q} +\vec{\Psi}(\vec{q},\tau)
\label{LagrangianMapping}
\end{equation}
While the solution $\Psi(\vec{q},\tau)$ describes the dynamics completely,
partial aspects of the dynamics may be described by other measures. 
The asymmetry of the fluctuation can be understood in terms of the 
eigenvectors and eigenvalues of the tidal tensor 
$T_{i,j} =\frac{\partial \Psi_i(\vec{q})}{\partial q_j}.$ This can be
visualized as an ellipsoid, which we shall refer to as the tidal ellipsoid,
 with principal axes along the eigenvectors of the tidal tensor with sizes
equal to the eigenvalues. For a spherically symmetric 
fluctuation, these eigenvalues are equal, while 
the departure from spherical symmetry may be characterized by 
different choices of functions of ordered eigenvalues of the tidal tensor. 
(See Appendix~\ref{OtherParametrizations} for some other popular 
choices in the literature.)
This was recognized and used in correcting for
ellipsoidal collapse of halos rather than spherical collapse in 
Press-Schechter like estimates of the mass function of dark matter halos 
~\citep{2001MNRAS.323....1S,2002MNRAS.329...61S,2001ApJ...555...83C}. 
From a theoretical side, we can describe the evolution of the 
distribution of these eigenvalues. Therefore, it is these dynamical 
quantities that we are interested in, even though they are not directly
observable. 

We will next proceed to describe 
observable quantities which relate to the shape of the voids, and then show
how functions of those observables trace functions of these dynamical 
quantities. Since voids form around 
minima of the density fields where the gradient of field vanishes, 
one can approximate the density profiles around the minima by truncating 
the Taylor expansion at second order. This gives density profiles that are
ellipsoidal in shape. One may expect voids to 
inherit this shape, and therefore be approximately ellipsoidal. 
In fact, 
voids have often been modeled as spherical 
(eg.~\citep{2004ogci.conf...58V}), while others have 
argued that the shapes of larger voids fit ellipsoids well only for 
smaller voids~\citep{2006MNRAS.367.1629S}. 
For irregularly shaped voids (obtained by suitable void identification
 algorithms), one can define a void ellipsoid by fitting a moment of 
inertia tensor to 
the positions of observed void galaxies $\vec{x}$ in Eulerian coordinates 
relative to the void center $\vec{x}^v$
$$S_{ij} = \frac{\sum_k (x^k_i - x^v_i)(x^k_j-x^v_j)}{\mathcal{N}},$$ 
where the index k runs over the observed galaxies in the void region, and 
$\mathcal{N}$ is the number of galaxies fitted. 
The void ellipsoid can be defined as the ellipsoid with principal axes 
along the eigenvectors of this mass tensor, and lengths proportional to
the square root of the eigenvalues $\{J_1,J_2,J_3\}$. 
Here, we shall ignore the 
discrepancy between the actual shape and this void ellipsoid.
Following ~\citet{2007PhRvL..98h1301P} (see
Appendix.~C of \citet{2009arXiv0906.4101L} for a calculation to first 
order), one can relate the eigenvalues of the tidal tensor 
$\{\lambda_1,\lambda_2,\lambda_3\}$ to the functions of 
the ratio of eigenvalues of the void ellipsoid which were called 
ellipticity.
Accordingly, 
the ellipticities $\{\epsilon,\omega\}$
 of the void ellipsoid are to first order
\begin{equation}
\label{eqn:VoidEllipsoidEllipticity}
\epsilon = 1 - \left(\frac{J_1}{J_3}\right)^{1/4} 
\approx 1 -\left(\frac{1- \lambda_1}{1-\lambda_3}\right)^{1/2},
\qquad
\omega = 1 - \left(\frac{J_2}{J_3}\right)^{1/4} 
\approx 1 -\left(\frac{1- \lambda_2}{1-\lambda_3}\right)^{1/2}.
\end{equation}
Clearly, this relation will be affected, at least to some extent, 
by more detailed dynamics. This would lead to $\epsilon$ measured from
data sets on voids being correlated with the functions of $\{\lambda_i\}$ 
with some scatter.
In computing parameter constraints, we 
shall account for this in terms of a variance in the quantity 
$\epsilon$ which also contains contributions from observational errors. 
We shall assess the 
impact of this assumption of the void shapes being perfect tracers of the 
eigenvalues by 
studying the degradation of constraints on increasing the variance in 
our study of systematics in Section.~\ref{ResultsIssues}.\\
\subsection{Distribution of Initial Eigenvalues of the Tidal Tensor}
\label{distribution_eigenvalues}
An observed void evolves from a fluctuation of low underdensity at early times
when the distribution of fluctuations was Gaussian. Given a void of 
a given density contrast, at a particular redshift, we wish to calculate
the distribution of eigenvalues of the tidal tensor of the initial 
fluctuation. 

At early times, the fluctuations are small enough, their growth can
be described by linear perturbation theory, and the distribution remains 
Gaussian. 
One can use the statistical properties of filtered isotropic and homogeneous Gaussian
fields to derive a probability distribution of the ordered eigenvalues 
of the tidal tensor given by the Doroshkevich formula. 
\begin{equation}
\label{DKV}
P(\lambda_1, \lambda_2, \lambda_3\vert \sigma_R ) =
	\frac{3375}{8\sqrt{5}\sigma_{R}^6} 
	\exp\left(-\frac{-3K_1^2}{2\sigma^2_{R}} + \frac{15 K_2}{2\sigma^2_{R}}\right)
	K_3
\end{equation}
where  $K_1 =\lambda_1 +\lambda_2+\lambda_3,\quad K_2 =\lambda_1\lambda_2 +\lambda_2\lambda_3+\lambda_3\lambda_1$, while $K_3 =-(\lambda_1-\lambda_2)(\lambda_2-\lambda_3)(\lambda_3-\lambda_1),$ and $\sigma^2_R$ is the 
variance of the smoothed overdensity field at the filtering scale $R$ at that time.
Note, that this gives the distribution of the size of the eigenvalues over
all spatial points. This distribution is extremely similar but slightly
different if restricted to the maxima of the Gaussian field 
~\citep{1986ApJ...304...15B}, or the minima of the Gaussian
field ~\citep{2009arXiv0906.4101L} which should 
evolve to voids. For the small fluctuations, one can use the 
Jacobian of the transformation from Eulerian to Lagrangian coordinates to
show that the sum of the eigenvalues $K_1$ can be identified with the 
density contrast.\\ 

It should be noted that this distribution depends on the filtering
scale $R_{\text{Smooth}}$ as a parameter while the size of voids is not important.
 This is appropriate for comparison with a dataset of voids obtained from 
redshift surveys by means of an algorithm which uses a filtering scale as 
a  parameter, rather than the void size. This is true for a class of
algorithms that define voids as regions of space where the smoothed matter 
density is a minimum  (eg.\citep{2007MNRAS.375..489H,2009arXiv0906.4101L}) 
with the smoothing scale $R_{	\text{Smooth}}$  being a parameter, with 
the actual size of voids not being crucial to the definition. 
On the other hand there are Void Finding algorithms which 
define voids as the largest contiguous underdense regions, obtained by some
 form of clustering algorithms. 
A corresponding parameter here is the size $R$ of the voids 
related to the void volume by $R^3\equiv \frac{3 V}{4\pi},$ while the 
smoothing scale is not crucial. 
While each algorithm might yield slightly different 
properties of voids, it would be expected that they are not too different. 
In Appendix~\ref{appendix:gesf}, we show that  a calculation based on the generalized excursion set formalism
can be used to calculate the distribution of eigenvalues of an initial
fluctuation that evolves to form a void of size $R$. The result of this
calculation supports the above result.  
\subsection{Evolution of the Tidal Eigenvalues}
\label{evolution_eigenvalues}
At low redshifts, gravitational collapse introduces non-linearities into
the evolution leading to non-Gaussian distributions of the density field. 
Thus, the distribution of the tidal eigenvalues of the previous subsection
 which assumed Gaussianity are not directly applicable. We study 
the evolution of these eigenvalues with time in two different methods, one 
based on the Zeldovich approximation and one based on \citet{1996ApJS..103....1B}.
\\

It is well known that non-linearity is manifested much less in
the displacement field or the gravitational (and the related displacement) 
potential than in the density field. Therefore before shell-crossing, the 
evolution of structures from  initial condition may be described by the 
Zeldovich approximation, where the displacement field is assumed to
be separable into a time dependent and time independent part.
$\Psi(q,\tau) = {D(\tau)\over D(\tau_0} \Psi(q,\tau_0)$,
 where $D(\tau)$ is the linear growth function. 
Hence, at a particular spatial point, its
eigenvalues $\lambda_i(\tau)$ at time $\tau$ evolve linearly from the 
eigenvalues $\lambda_i(\tau_0)$ 
at some initial time $\tau_0$ as 
$\lambda_i (\tau)= D(\tau)\lambda_i(\tau_0)/D(\tau_0).$  
Rewriting the early time eigenvalues in the Doroshkevich formula 
(Eqn.~\ref{DKV}) in terms of the eigenvalues at time $\tau,$ 
one can then find a distribution of eigenvalues at
any time to be given by the Doroshkevich formula where the $\sigma_R$ is 
replaced by $D(\tau)\sigma_{R}/D(\tau_0)$, the linearly 
extrapolated variance over the Lagrangian smoothing scale $R$.
The formula is exactly the same as Eqn.~\ref{DKV} with the variance 
$\sigma^2_R$ being replaced by the linearly extrapolated variance
$\sigma^2(R,z)$, and $\lambda_i$ replaced by the eigenvalues at the 
redshift of the void. 
Further, since the sum of the eigenvalues $K_1$ at early times was equal
to the density contrast at that time, the term $K_1$ is equal to the 
linearized density contrast of the time of the void
\begin{equation}
\delta_{lin}(\tau) = \frac{D(\tau)}{D(\tau_0)}\delta(\tau_0) =\frac{D(\tau)}{D(\tau_0)} 
(\lambda_1(\tau_0)+\lambda_2(\tau_0)+\lambda_3(\tau_0))
=(\lambda_1(\tau)+\lambda_2(\tau)+\lambda_3(\tau))
\label{eqn:deltalin}
\end{equation}

In regions of high density peaks where structure forms, it has been found
that modeling the density growth as a collapse of a homogeneous 
ellipsoid leads to a better approximation to N body simulations. It is 
unclear whether this should also be true for low density regions like 
voids. In Appendix~\ref{appendix:EllipsoidalCollapse}, we study the 
evolution of the eigenvalues of the tidal tensor  
based on ellipsoidal collapse ~\citep{1996ApJS..103....1B} 
and find the differences with the evolution computed using Zeldovich 
approximation to be small.


\subsection{Cosmology Dependence of the Distribution of Ellipticity}

Therefore, using the Zeldovich approximation, one can write down the probability distribution of the eigenvalues of the tidal tensor at any time. 
Further, using the relations of the ellipticities of the 
void (Eqn. ~\ref{eqn:VoidEllipsoidEllipticity}) and the relation of the linearly 
extrapolated  
density contrast to the eigenvalues $\{\lambda_1,\lambda_2,\lambda_3\}$, 
one can recast this as the joint distribution of the ellipticities 
$\{\epsilon,\omega\}$ given the smoothing scale and the linearly 
extrapolated density contrast. Following Park and Lee, we define $\mu,\nu$ and 
write the probability distribution for the larger ellipticity $\epsilon$ 
\begin{eqnarray}
\label{eqn:ellipticitydist}
\mu & =  &\left(J_2/J_3\right)^{1/4},  \qquad 
\qquad 
\nu  =  \left(J_1/J_3\right)^{1/4}  \nonumber\\
 P(\mu, \nu \vert \sigma_{lin} (R,z), \delta_{lin}(z)) &=&
\frac{3^4/4}{\Gamma(5/2)}\left(\frac{5}{2~\sigma^2_{lin}(R,z)}\right)^{5/2}
\exp
\left(
	-\frac{5 \delta_{lin}^2(z)}{2~\sigma_{lin}^2(R,z)} +
	\frac{15 K^{\delta}_{2}}{2~\sigma_{lin}^2(R,z)}
\right)
K^{\delta}_3 J\nonumber\\
P(\epsilon \vert \sigma_{lin}(R,z),\delta_{lin}(z)) & = &
\int_{1-\epsilon}^1 d\mu P(\mu,1-\epsilon \vert \sigma_{lin} (R,z), \delta_{lin}(z))
\end{eqnarray}
where $K^{\delta}_{2},K^{\delta}_{3}$ are the values of 
$K_2,K_3$ in Eqn.~\ref{DKV} in terms of $\mu,\nu$ when the 
constraint of Eqn.~\ref{eqn:deltalin} holds, and $J$ is the Jacobian in 
the transformation from the coordinates $\{\lambda_1,\lambda_2,\delta_{lin}\}$
to $\{\mu,\nu,\delta_{lin}\}.$
This last equation gives the probability distribution of the larger 
ellipticity $\epsilon$  marginalized over the smaller ellipticity 
$\omega$. 
It depends on the cosmology only through the 
linearly extrapolated variance $\sigma^2_{lin}(R,z)$ of density 
fluctuations 
$\delta(x,z)$ smoothed at a certain filtering scale $R$ by a window function 
$W_R(x,x^\prime).$
\begin{equation}
\sigma^2_{lin}(R,z)\equiv \lept{\delta^\star_{R}(x,z)\delta_{R}(x,z)}
=D^2(\tau)\sigma^2_{R} 
\qquad
\delta_{R}(x,z) =\int d^3x^\prime \delta (x,z) W_{R}(x,x^\prime)
\end{equation}
where $D(\tau)$ is the growth function and $\sigma_R$ is evaluated at 
early times. 
For qualitative understanding, 
it is useful to think of the variance depending on cosmology through 
$\sigma_R$ which depends on the primordial power spectrum and the 
wave mode dependent transfer function, and the subsequent 
scale independent growth described by the growth function $D(\tau)$. While
the transfer function depends on most of the cosmological parameters, in 
most models dark energy does not become significant at early times. 
Therefore most of the effects of dark energy are embedded in the growth 
function. Closed analytic forms for the growth function are not 
known for non-flat cosmologies, with time varying equations of state dark 
energies, but~\citet{2005A&A...443..819P} 
improves upon a fit to the growth function by~\citet{2003ApJ...590..636B}, 
so that the fit works for non-flat cosmologies having dark energy with 
time varying equation of state as long as they are close to flat LCDM 
models, even when the equation of state is less than -1. If we consider the
CPL parametrization
\begin{equation}
w(z) = w_0 + w_a \frac{z}{z+1},~
\end{equation}
we see in the left panel of Fig.~\ref{GrowthDerivs} that
the growth function changes more dramatically as a function of $w_0$ than 
for $w_a$ for redshifts below unity. Thus, we expect, that constraints 
from  voids in these redshift ranges should be stronger on $w_0$ than on $w_a$.
From the right panel of Fig.~\ref{GrowthDerivs}, 
we can see the effect of the filtering scale 
$R$ on the distribution. Since $\sigma_{lin}(R,z)$ is a monotonically 
decreasing function of $R,$ a larger filtering scale (a) shifts the 
distribution towards smaller values of $\epsilon,$ and (b) sharpens the 
distribution. This is consistent with intuition based on previous studies
~\citep{1979ApJ...231....1W,1984MNRAS.206P...1I,1996MNRAS.281...84V}. 
Leaving all other variables the same, increasing $R$ corresponds to 
excluding the smaller voids. 
Since the variation of possible values is caused by the variance 
in the Gaussian distribution, a smaller value of $\sigma_{lin}(R,z)$ also
corresponds to a sharper distribution.\\

In this paper, we shall assume that all voids are found at a linearized 
density contrast of $\delta_{lin} =-2.81,$ the underdensity at shell 
crossing. 
We shall compute $\sigma_{lin}(R,z)$ directly from numerical 
integration of the smoothed density fluctuations evolved by a modified 
version of the Boltzmann code \verb CAMB ~\citep{2000ApJ...538..473L}. 
\begin{figure*}[!htp]
\begin{center}
\includegraphics[width=0.9\textwidth]{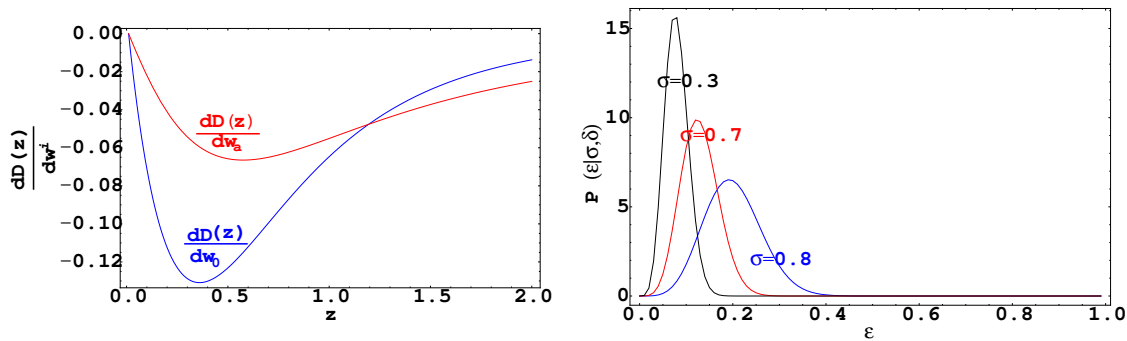}
\caption{Left Panel:The Derivative of the growth function with respect to 
the dark energy parameters $w_0$, and $w_a$. The growth function shown has
 been normalized to unity at a redshift of 0.01. Right Panel: The theoretical distribution of the largest ellipticity $\epsilon$ as a function of 
$\sigma(R,z)$ for $\delta_{lin}$=-2.81}
\label{GrowthDerivs}
\end{center}
\end{figure*}
\section{Distribution of Ellipticity: 
Connecting to Observations}
\subsection{Estimate of Voids to be found from a survey}
Next, we proceed to estimate the number of voids that we expect to find in 
a certain survey. We model a survey by considering a redshift survey, 
which can measure the redshifts of the galaxies  up to a limiting visual 
magnitude of $m_L$ in a given filter and from a minimum redshift of 
$z_{min}$ to a maximum of $z_{max}$. 
In case of photometric surveys, the errors
in redshift can be much larger, leading to errors in the size of the ellipse along the 
line of sight, consequently the distribution of ellipticities will have 
to be marginalized over this error. Here, we will limit our considerations
to spectroscopic surveys, where the error in measuring the redshift of 
the galaxies $\sim 10^{-4}$ is negligible. 

In order to estimate the number
of voids of a particular size at a particular redshift, we use the 
Press-Schechter formalism to determine the number density of voids in a 
redshift bin centered at $z,$ with Eulerian comoving radius between $R_E$
and $R_E + d R_E.$ Simulations indicate that the number density of voids
peaks at a density contrast of $\delta  \approx -0.85$~\citep{2007PhRvL..98h1301P}, we shall consider all the voids to have 
a density contrast of 0.8, which can be seen to correspond 
to a linearly extrapolated density contrast of -2.81 using the fitting 
function in ~\citet{1996MNRAS.282..347M}. While the usual Press-Schechter 
formalism matches simulations well at redshift ranges below $\approx 2,$
it fails to predict the number of voids correctly at small scales due to
the `void in cloud problem' , which can be avoided if at  
each redshift, we restrict ourselves to scales larger than the 
non-linearity length scale (Lagrangian) $R_{min}^{VinC}(z)$ where 
$\sigma (R_{min}^{VinC}(z),z)= 1$~\citep{2004MNRAS.350..517S}. Then, the 
Press-Schechter formalism reliably predicts the number of voids with the 
replacement 
$\delta_c = 1.69 \rightarrow \delta_v =-2.81$ in the standard 
Press-Schechter formalism ~\citep{1974ApJ...187..425P}.
The number of voids of a particular size can then be found by integrating 
over the cosmological volume in the redshift bin, and over the range of 
radii allowed. 
\begin{eqnarray}
n_v(R_E,z) dR_E~&=&~ 
\frac{3 }{2\pi R_E^3}
P\left(-\frac{\vert\delta_v(z)\vert}{\sigma_{R_E}}\right)
\left
\vert
\frac{d}{dR_E}\frac{-\vert\delta_v (z)\vert}{\sigma_{R_E}}
\right
\vert
dR_E \\
N_{void}~&=&~
\int_z^{z+\Delta z} d\Omega 
dz \int_{R_E}^{R_E+\Delta R_E} d R_E \;  
\frac{dV}{dz d\Omega} \; n_v(R_E)\nonumber
\end{eqnarray}
where $P(y) = \sqrt{\frac{1}{2}}exp(-y^2/2).$
The number density of voids thus depends exponentially on $\sigma_R$
and therefore the number of voids is extremely sensitive to the minimum 
radius used. Since voids are detected by observing 
galaxies rather 
than the matter density, the number of voids detected 
with small radii will be strongly affected by shot noise 
(discussed in subsection \ref{EffectsOfShotNoise}). We therefore only consider voids 
with radii greater than a critical radius $R \geq R_{\mbox{min}}^{\mbox{shot}}(\mbox{z,Survey})$.
For our purposes then, the minimum of the range of radii of voids 
at a redshift $z$ considered must be set to the maximum of 
$R_{min}^{Vinc}(z)$ and $R_{min}^{shot}(z,Survey).$\\
 
We now explain our method for computing  
$R_{min}^{shot}(z,Survey),$ from the parameters for a survey. 
The minimum radius of voids that we will consider should be
related to the average separation of galaxies \emph{observed} 
$l_{\mbox{sep}}(z)$ at the redshift $z$ by the survey in question. 
We choose this relationship to be linear 
$R_{\mbox{min}}^{\mbox{shot}}(\mbox{z,Survey})= A l_{\mbox{sep}}(z),$ 
and relate the average separation to the average number density of 
observed galaxies $n^{bg}_{gal}(z)$ at that redshift for the survey. A 
choice of $A=2$ implies that the probability that a detected void is just
due to shot noise is less than 0.5 percent while such a scenario for $A=1$
is of the order of 50 percent, though void identification algorithms can 
do better, since they can exploit the contrast between voids and their 
higher density environments. In any case, the interesting regime is in 
between these numbers and we shall later explore the sensitivity of 
constraints to this range.

This background number density of observed galaxies $n^{bg}_{gal}(z)$ can be 
related  to the survey parameters. 
The mean number density of galaxies in the background universe can be 
calculated from the luminosity function 
~\citep{2001AJ....121.2358B} 
of galaxies at the filter band
used in the survey by,
\begin{equation}
\label{nbggal}
n^{bg}_{gal}(z)=\int_{-\infty}^{M_L} dM \Phi_X(M,z)
\end{equation}
where $\Phi_X$ is the luminosity function for the filter $X$ and $M_L$ is 
the limiting absolute magnitude of objects at redshift $z$ which
are observed by the survey. 
It can be calculated from the limiting apparent magnitude of the survey 
$m_L$ by using the formula,
\begin{equation}
M_L= m_L  -5 \log_{10}D_L(z) +5 - A(z) - K(z)
\end{equation}
Here $D_L(z)$ is the luminosity distance to the redshift $z$ in units of 
pc, $A(z)$ is the correction due to extinction and $K(z)$ is the 
K correction 
arising from the difference in the observed luminosity of and the rest 
frame luminosity of an object in a particular frequency band due to 
redshifting of photons.
  
We note that $R_{min}^{VinC}$ depends on the cosmology, but is independent
of the survey, while $R_{min}^{shot}(z,survey)$ also depends on the survey
through the filter band, and the limiting magnitude. 
A plot of $R_{min}^{noise}$ and $R_{min}^{VinC}$ for surveys considered in 
this paper is shown in Fig.~\ref{Fig:PlotRmin}. Thus, our estimate of the number of voids 
identified by each survey depends on the cosmology, 
the value of the proportionality constant $A$ and the survey parameters.
\begin{figure*}[!hpt]
\begin{center}
\includegraphics[width=0.45\textwidth]{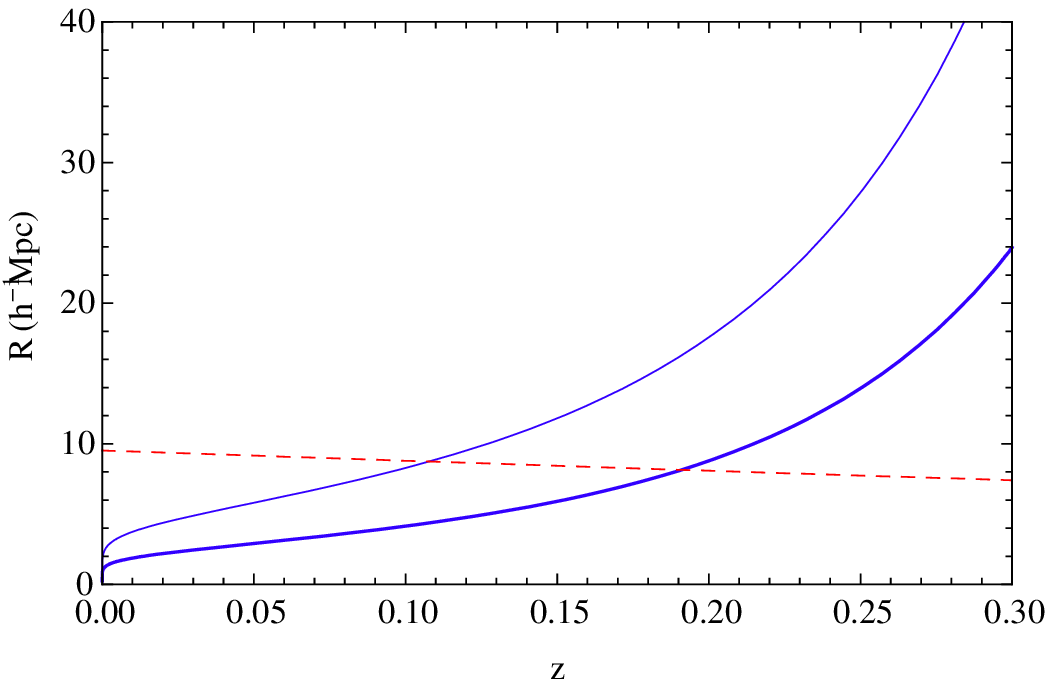}
\includegraphics[width=0.45\textwidth]{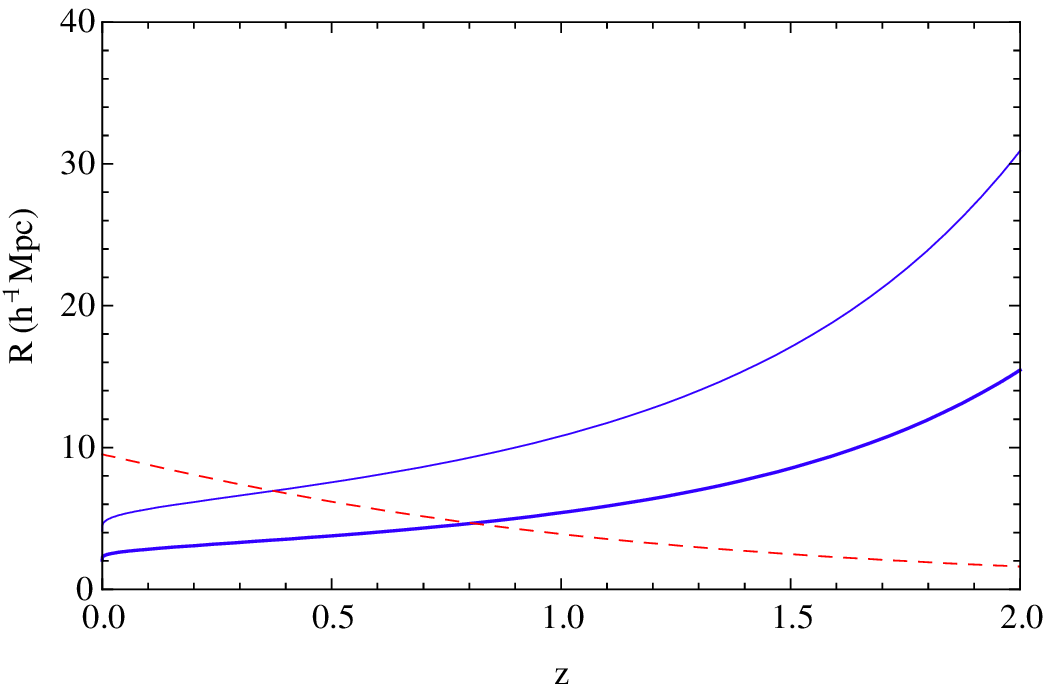}
\caption{Setting the minimum size of voids: the dashed red curve shows the
$R_{min}^{VinC}$, while the solid thin (thick) curves show the (twice) the average separation of observed galaxies for a SDSS DR7 like survey (left) and a EUCLID like survey (right).
At a particular redshift, we only consider
voids with sizes larger than both these scales.}
\label{Fig:PlotRmin}
\end{center}
\end{figure*}

\begin{table*}
\label{VoidSurveys}
\caption{Surveys and parameters used for estimating the number of voids
that can be found by the survey. We chose a survey like SDSS DR7
as an example of a current survey, and EUCLID as an example of 
a futuristic survey. For reference, we show the number of galaxies that 
these surveys are expected to observe.}
\begin{tabular}{|c|c|c|c|c|c|}
\hline
Survey & $f_{sky}$ & Freq Band & Limiting Magnitude & Number of Voids &
Number of Galaxies\\
	&	& 	&	& $A=2$,$A=1$ & \\
\hline
\verb=SDSS= DR7\footnote{\url{http://www.sdss.org/dr7/coverage/index.html}} & 0.24 & r  & 18  &   1292,3104 & 1.7 $10^6$\\
\hline
\verb=EUCLID= \footnote{\url{http://hetdex.org/other_projects/euclid.php}} & 0.48 & K & 22  &   1.4 $10^5$, 2.3 $10^6$& 5.2 $10^8$ \\
\hline
\end{tabular}
\end{table*}
\section{Results}
\subsection{Likelihood function and Fisher matrix}
In order to study the potential constraints on cosmological parameters,
we need to write down a simple model for the data. We assume that by 
applying appropriate simulation algorithms, we can identify a set of voids 
at each redshift bin corresponding to a particular smoothing scale. We 
expect to measure the ellipticities of each of these voids with some
error. We model the error as an additive Gaussian noise $n$ on the 
ellipticity $\epsilon_s$:
\begin{equation}
\epsilon_d (R,z) = \epsilon_s(R,z) + n, \quad n \sim G(0,\sigma_\epsilon )
\end{equation}
$\epsilon_s$ itself is a random variable following the distribution of
the ellipticities at the relevant redshift. Then we can write down the 
likelihood function, which is the probability for finding a void with 
a measured largest ellipticity $\epsilon_d$ given the cosmological 
parameters
\begin{equation}
L(\epsilon_d \vert \Theta ) = 
\int d\epsilon_s P(\epsilon_d \vert \epsilon_s) 
P(\epsilon_s \vert \sigma_\epsilon ,\Theta)
\label{LikeIndividual}
\end{equation}
One expects that the 
error in measuring the ellipticities will be set by the errors in measuring
the principal axes of the void ellipsoid. For a spectroscopic survey, the 
positions of galaxies are well measured. Ignoring effects of redshift 
distortion/finger of god effects the precision level of the measurement 
of the principal axes would be set by the errors in the void finding 
algorithm. Of course, this will be limited by the relative sizes of the 
void wall thickness to the void radius $\Delta$. For $\Delta \sim0.1-0.4$,
$\epsilon \approx 0.2$ around the maximum for standard cosmological 
parameters, the error in $\epsilon$ is of the order of 0.1.
The errors in the measurement of each void is statistically independent.
Thus the likelihood function for an entire data set 
consisting of voids at different redshifts can be computed as the 
product of Eqn. \ref{LikeIndividual} for each void. Consequently, the log of 
the likelihood function ${\cal{L}}(\epsilon_d \vert \Theta )$ is additive
for each void.

Given the likelihood function for a single void, 
one can compute the Fisher matrix F defined as an expectation over 
all possible sets of data,
 \begin{equation}
F_{ij}=
\left\langle
\frac{\partial{\cal{L}}(\epsilon_d \vert \Theta, \sigma_\epsilon)}
{\partial \Theta_i}
\frac{\partial{\cal{L}}(\epsilon_d \vert \Theta, \sigma_\epsilon)}
{\partial \Theta_j}
\right\rangle
= 
\int_0^1 d\epsilon_d L(\epsilon_d \vert \Theta, \sigma\epsilon)
\frac{\partial{\cal{L}}(\epsilon_d \vert \Theta, \sigma_\epsilon)}
{\partial \Theta_i}
\frac{\partial{\cal{L}}(\epsilon_d \vert \Theta, \sigma_\epsilon)}
{\partial \Theta_j}
\end{equation}
where all the derivatives are taken at a fiducial choice of the 
cosmological parameters $\Theta_p$. 
Since, in our model the error in measuring the ellipticity is independent
of the cosmological parameters, and the ellipticity depends on the 
cosmological parameters through the variance of the fluctuations 
$\sigma^2_R$ only, we can factorize this into a matrix of 
mixed partial derivatives of $\sigma_R$ with respect to the cosmological 
parameters, and the derivatives of the log likelihood with respect to 
$\sigma_R$. We evaluate both of these derivatives numerically. The 
main contribution to the derivatives comes from the regions where the 
probability is smallest. However, these contributions are suppressed in the
expectation values, since these regions have low probabilities.
Finally, we must sum this contribution for the Fisher matrix 
over all the voids in the data set. The result thus depends critically 
on the number of voids in the data set. 

\subsection{Forecasts of constraints on the CPL parameters}
We consider Fisher forecasts for a cosmology with the 
non-baryonic matter assumed to be cold, neglect effects of neutrino masses
and parametrize the evolution of the dark energy equation of state with 
a CPL parametrization. The primordial perturbations are assumed to be 
Gaussian distributed, and characterized by a spectrum which is a power law
with an initial amplitude $A_s$, and a scale independent tilt $n_s$. 
The distribution of ellipticities depends on both the amplitude of 
primordial perturbations, and the spectral index through the 
dependence of the variance on the scale of smoothing. As is well known, 
these quantities $A_s, n_s$ are not exactly known, and have a degeneracy 
with $\tau,$ the optical depth of reionization. Further, the constraints 
on the equation of state parameters can depend strongly on the knowledge of
the curvature parameter ~\citep{2009arXiv0903.2532B}.
We therefore consider forecasts for constraints on the CPL parameters 
$w_0,w_a$ after marginalizing over all other cosmological parameters from 
a maximal set shown in Table.~\ref{Params}, 
along with the fiducial values used for computing the Fisher forecasts.  

All of these parameters are not well constrained by a single experiment.
Consequently, we shall consider Fisher forecasts using ellipticity 
distribution of voids from two spectroscopic surveys: the 
recent \verb SDSS  DR7 and the futuristic~\verb EUCLID 	with the survey 
parameters assumed summarized in 
Table.~\ref{VoidSurveys}. We will assume 
$A =1,$ 
$\sigma_\epsilon=0.1 .$ Following the work in ~\citep{2009arXiv0906.4101L},
 we will identify the smoothing scale as being a quarter of the radius of 
the void. 
For CMB constraints, we will consider Fisher
forecasts computed from \verb PLANCK ~    
\footnote{\url{http://www.rssd.esa.int/index.php?project=Planck}}
The expressions for the Fisher matrix for CMB data are given in 
~\citet{1998astro.ph..4168T}. 
The survey parameters for \verb PLANCK ~are taken from the Table. 1.1 
of the PLANCK~Bluebook~\citep{2006astro.ph..4069T}, 
and are summarized in Table.~\ref{PlanckParams}. 
We consider Fisher forecasts of Supernovae from two surveys: for a survey 
like Dark Energy Survey the number of supernovae expected is of the order 
of 1300, and the maximum redshift is around 0.7. We model this with a 
redshift distribution taken from~\citep{2008ApJ...675L...1Z} designed to 
be cut off at z=0.7, and assume perfect measurement of redshift, due to
plans of spectroscopic follow-up. The errors in the 
magnitude are assumed to be of the order of the intrinsic dispersion from
light curve fitting techniques today (0.15). We also consider a futuristic 
photometric Supernova IA survey \verb LSST ~\citep{2009arXiv0912.0201L},
where about 500,000 SNe IA suitable for constraining dark energy 
parameters could be observed. We model the errors by assuming magnitude 
errors of the order of 0.12 from intrinsic dispersion, and photometric 
errors in redshift determination of the order of 
$\Delta z=0.01(1+z),$ and assuming that this adds an error 
$\frac{dm}{dz} \Delta z$ in quadrature to the intrinsic dispersion. 
We use the redshift distribution in Table 1.2 of the~\citep{2009arXiv0912.0201L}
to model the redshift distribution of the LSST survey.
\begin{table*}
\label{Params}
\caption{Parametrization of the cosmology and the fiducial values chosen
for the maximal set of parameters used in evaluating the Fisher forecasts.
Constraints are also discussed after imposing flatness.}
\begin{center}
\begin{tabular}{|c|c|c|c|c|c|c|c|c|}
\hline
$\Omega_{\rm{b}}\rm{h}^2$ & $\Omega_{\rm{c}}\rm{h}^2$
& $\theta$ & $\tau$ & $\Omega_{\rm{k}}$ & $w_0$ &$w_a$ & $n_s$ &
 $log(10^{10}A_s)$ \\
\hline
 0.02236 & 0.105 & 1.04 & 0.09 & 0.0 & -1 & 0 & 0.95 & 3.13 \\
\hline
\end{tabular}
\end{center}
\label{PriorsTable}
\end{table*}

\begin{table*}
\caption{Parameters of the PLANCK~Survey 
used in determining CMB constraints}
\label{PlanckParams}
\begin{tabular}{|c|c|c|c|c|c|c|c|}
\hline
Frequency Channel 
& 30 & 44 & 70 & 100 & 143 & 217 & 353\\
(GHz)  &  & & & & & & \\
\hline
Beam Width & 33 & 24.0&  14.0 & 10.0 & 7.1&  5.0 & 5.0 \\
(FWHM) arc min     & & & & & & & \\
\hline
Temperature Noise per Pixel &2.0 & 2.7 & 4.7 & 2.5 & 2.2 & 4.8  & 14.7 \\
\hline
Polarization Noise per Pixel & 2.8 & 3.9 & 6.7 & 4.0 & 4.2 & 9.8 & 29.8 \\
\hline
\end{tabular}
\end{table*}
\begin{figure*}[!hpt]
\begin{center}
\includegraphics[width=0.95\textwidth]{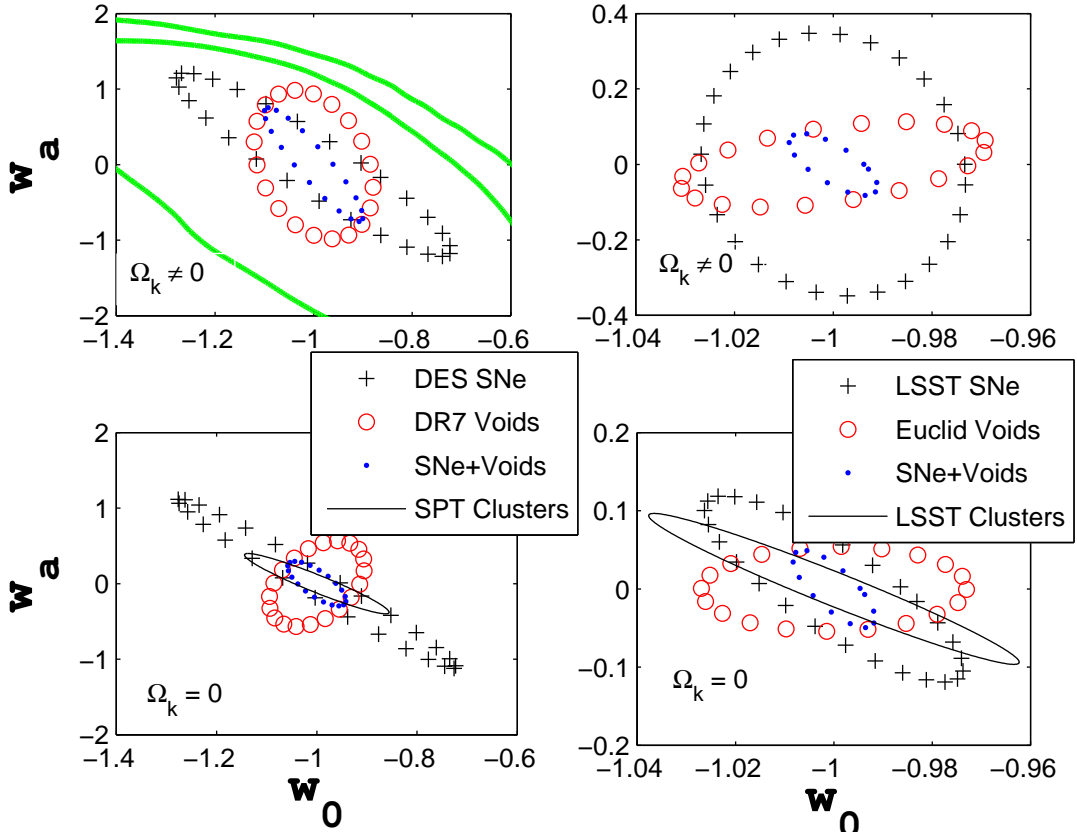}
\caption{~{\textbf{Comparison of forecasts on one $\sigma$ constraints on 
the CPL parameters with standard probes} using the identification $R_{	\text{Smooth}}=R_{Void}/4$, $A=1$, and $\sigma_\epsilon=0.1$:} for data from the near future (left panels) and futuristic data (right panels). PLANCK and HST priors were used 
in all of these forecasts. For reference, we show the current constraints
~\citep{2009arXiv0903.2532B} in the thick green contours, and 
forecasted constraints from clusters (number counts and power spectrum) + 
PLANCK taken from ~\citep{2004PhRvD..70l3008W}.
}
\label{Constraints}
\end{center}
\end{figure*}
In Fig.~\ref{Constraints}, we present the constraints on the equation of 
state parameters $w_0, w_a$ by combining constraints for two sets of data
(a) data representative of current or near future (left panels), 
and (b) data 
representative of more futuristic data (right panels). The forecasts
for one sigma constraints using void ellipticities + CMB + HST 
are shown in open circles, 
assuming $A=1.$
The error in measuring 
the ellipticities $\sigma_{\epsilon}$ is taken as 0.1. 
The ellipses made of black "+" show the 
constraints for SNe + HST +CMB (PLANCK). 
The solid, thick, blue ellipses show the 
constraints when these constraint are combined (CMB (PLANCK) 
+SNE +HST + Voids).
In the left panels of the figure, the voids considered
are from a survey like SDSS DR7, and the SNe considered are from a
survey like DES. In the right panels the voids considered are from a 
futuristic survey like EUCLID, and the SNe are from a futuristic 
photometric survey like LSST. The upper panels show the constraints 
marginalized over all other parameters in the maximal set, while the 
lower panels show the marginalized constraints for a flat universe. 
For reference, we show the thick, green contours showing the
one sigma constraints from current SNe (Union) + HST + CMB (WMAP 5) data
from~\citep{2009arXiv0903.2532B}. For the flat universe in the lower panels
, we also show the constraints from CMB (PLANCK) + HST + Clusters (Power
spectrum + Number counts) 
from ~\citep{2004PhRvD..70l3008W}. In the lower left panel the 
Clusters considered from the SPT survey, while the lower right panels
 show the constraints from clusters from LSST.

Firstly, these figures show that the inclusion of constraints from void 
ellipticities significantly improves parameter constraints and the 
constraints from Voids along with CMB and HST data are comparable to the 
joint constraints obtained by using Supernovae IA, CMB and HST data both in
the near future and the far future. As is common, following 
~\citet{2006astro.ph..9591A}, we quantify this in terms of a Figure of 
Merit (FoM) which is inversely proportional to the area of the two sigma 
contours 
(ie. proportional to the inverse of the determinant of the 
$w_0,w_a$ submatrix of the inverse of the Fisher Matrix). 
We calculate the FoM relative to the FoM without voids 
for each of the upper panels:
$$FoM(\rm{experiments}) = det(\rm{SNE+PLANCK+HST})/det(\rm{experiments})$$
where $\rm{experiments}$ refer to the combination of experiments we 
consider the FoM for, and the SNE experiments in the numerator refer to
the DES for the left panel, and LSST for the right panel. The relative
FoM for these results are shown in ~\ref{table:FoM} 
(for $A=1,\sigma = 0.1$). We see that the constraints with the 
use of (Voids + CMB + HST) is not good as, but somewhat comparable 
(Relative FoM =0.6) to the constraints due to (SNe + CMB + HST), but adding
the void constraints to the SNE +CMB +HST data offers a moderate gain 
(FoM = 13.3). For the futuristic case, the use of (Voids + CMB + HST) is 
is better than the corresponding (CMB + SNe +HST) data (FoM=70.4), while
combining these constraints improves the FoM by a factor of 2500.\\
    
We should stress that even the results for the SDSS DR7 survey 
(with a relative FoM of 0.6) are promising, because they are a
different way of probing the dynamics and therefore can be potentially
useful in determining consistency of the underlying cosmological 
model. Clearly the addition of void ellipticities as an observable for 
parameter estimation increases our knowledge of the cosmological 
parameters in other cases. 
\subsection{Study of Possible Systematics}
\label{ResultsIssues}
While we have shown that our forecasted constraints are extremely 
promising, we have used order of magnitude calculations often based on 
first order results in semi-analytic models. 
By doing N-body simulations of large scale structure it is possible to 
replace these by more accurate calculations, and use it for estimating 
cosmological parameters. This would be the goal of future work in this 
direction. But is it possible that when such a rigorous analysis is 
carried out the constraints might get terribly degraded and not be 
interesting any more? The objective of this subsection is to address this 
concern by trying to list the major assumptions that would need to be 
replaced in a rigorous calculation, and trying to obtain a sense for how 
far these constraints might be degraded. We discuss the basic assumptions
 and explain how we might expect these factors to affect the forecasts.\\

\begin{figure*}
\begin{center}
\includegraphics[width=0.95\textwidth]{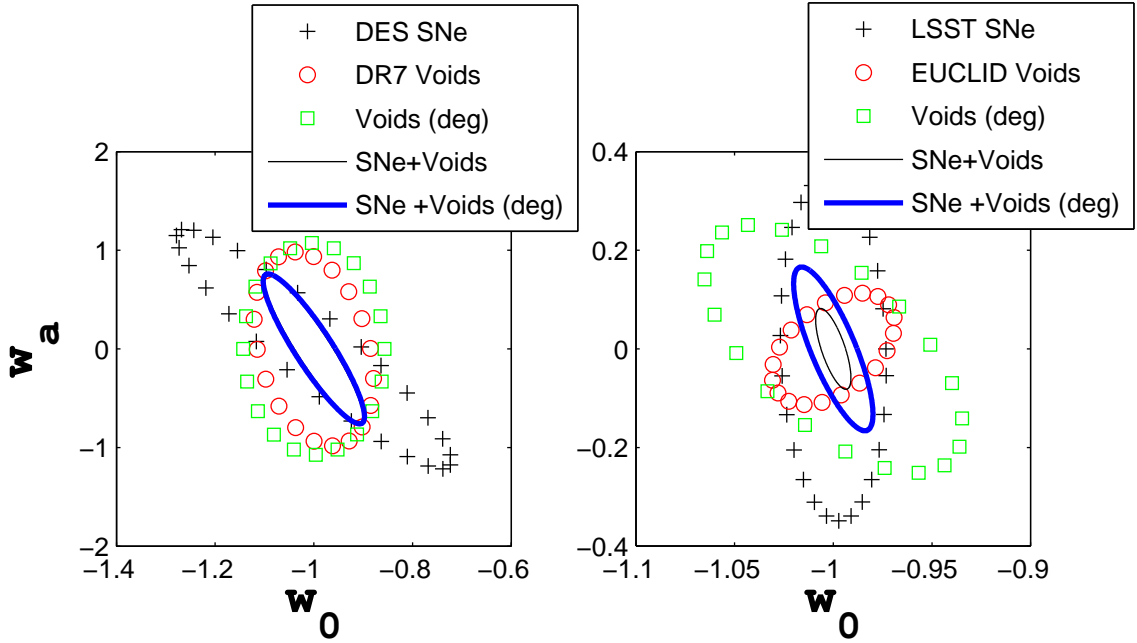}
\caption{~\textbf{Effects of Shot Noise: Sensitivity of one sigma 
constraints to the efficiency of the void finder:}
Degradation of constraints due to low efficiency of 
the void finder with the 
constraints shown in Fig.~\ref{Constraints} assuming high efficiency
from the near future (left panel) and futuristic data (right panel). 
HST and PLANCK constraints were used in all these plots.} 
\label{VariationWithDelta}
\end{center}
\end{figure*}
{(a)\it{Effects of Shot Noise on the Number estimate of Voids:}} 
\label{EffectsOfShotNoise}
Our constraints are obviously dependent on our 
estimate of the number of voids that would be detected in a particular 
survey. Thus, regions of space which are
not true voids but get misidentified as voids would cause a spurious 
enhancement of signal. Recall that voids have been defined as regions of
space where the \emph{total matter density} is low (or minimum) 
but are identified by the low density of \emph{galaxies} which are biased baryonic
tracers of the density field. The lack of direct knowledge of the dark
matter density field is often addressed in the context of the 
Poisson Sample Model, where density contrast of galaxies is described as 
a Poisson point process with a mean density proportional to the dark 
matter density. Thus, there is a chance of identifying a region which has 
low density of galaxies but not dark matter as a void. 
Consequently, due to shot noise, one can 
only confidently infer a region of low galaxy density to be a void if 
the region is large relative to the average separation 
$l_{sep}(z)\sim (n_{gal}^{bg})^{-1/3}$ of visible galaxies 
at that redshift. This means that small voids might not really be voids, 
and the problem is exacerbated by the fact that the number of voids 
increases exponentially with smaller sizes of voids.
A sophisticated treatment of this problem
would associate a probability to describe the confidence of detection 
(for example as in ~\citet{2008MNRAS.386.2101N}) and incorporate that in
the Likelihood. We use a rough model to estimate the importance of this 
effect by only choosing a minimum radius $R_{min}^{shot}(z,survey)$ 
of voids related to the $l_{sep}(z)$ as discussed before. 
A larger value of $A$ results in a larger values of 
$l_{sep}(z)$ which leads to a higher threshold for the minimum size of 
voids observed in the survey. 
Since the minimum radius of voids is set by the maximum of
this survey dependent $R_{min}^{shot}$ and the survey independent 
$R_{min}^{Vinc}$ (Void in Cloud), this changes the numbers of voids strongly where 
$R_{min}^{shot}$ is much smaller than $R_{min}^{Vinc}$.
We therefore compare the
constraints for a pessimistic value of $A =2$ to the constraints
obtained in Fig.~\ref{Constraints} with $A=1$. 
In Fig.~\ref{VariationWithDelta}, we show the Fisher forecasts for values of $\Delta$ assuming the same
value $\sigma_\epsilon=0.1$ for both cases. The red ellipse with
open circles show the constraints from Voids (SDSS) + HST + PLANCK  in the 
left panel, and Voids (EUCLID) + PLANCK + HST (right panel) for 
$A=1,$ while the open green squares show the same constraints if 
$A=2$. When additionally, supernovae data is used: 
on the left panel we have DES SNe + HST +PLANCK + SDSS Voids, while on the
right panel we use LSST SNe + HST + PLANCK + EUCLID Voids. The solid, thin 
black ellipse shows these constraints for $A=1$, while the solid 
thick blue ellipse show these constraints for $A=2$. 
For reference, we use the black "+" to show the constraints from  
DES SNe+ PLANCK + HST on the 
left panel, and LSST SNe + PLANCK + HST on the right panel. 
Clearly, while
the constraints change, there is no severe degradation due to shot noise
for the case based on DR7 survey, while this is somewhat important for 
the case based on EUCLID.
We 
summarize the degradation in terms of a relative FoM in 
Table.~\ref{table:FoM}.
\begin{figure*}[!tp]
\begin{center}
\includegraphics[width=0.9\textwidth]{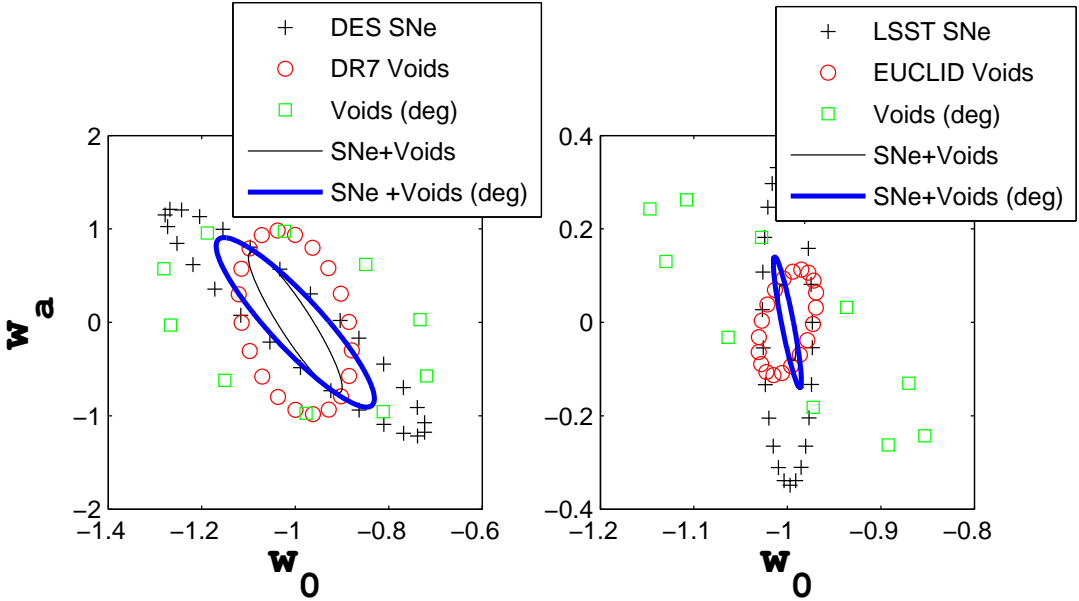}
\caption{{\textbf{Impact of Bias}}: Degraded constraints on voids due to 
marginalization over a linear scale independent bias compared to 
constraints shown in Fig.~\ref{Constraints} for data from the near future 
(left panel) and futuristic data (right panel). HST and PLANCK priors were
considered for all of these plots.
}
\label{BiasFig}
\end{center}
\end{figure*}

{(b) \it{Bias:}} Since the observations pertain to galaxies rather than 
the dark matter distribution, we have no direct knowledge of the dark 
matter distribution even though the galaxy distribution and dark matter 
distribution are correlated. The qualitative understanding of the situation
is that galaxies form due to the collapse of baryons into gravitational 
potential wells of collapsed dissipation-less dark matter. The simplest 
 popular idea of linear scale independent bias models this by assuming that
locally, the dark matter density contrast $\delta_g$ is proportional to the
the total matter density contrast $\delta_m$, and the constant of 
proportionality is called the bias $b$. Bias different from unity affects
our forecasts in two ways: (i) first, the Lagrangian radius of the void 
is estimated incorrectly as a function of $\delta_g$ rather than 
$\delta_m$. This leads to the use of a variance $\sigma_R$ on the incorrect scale, 
and second (ii) since we use the probability distribution of the 
eigenvalues conditioned on the density contrast of the voids, this changes
the distribution of the eigenvalues. 
To address the issue of bias, we recalculate the forecasts by adding an 
extra  parameter, the bias $b$ to our set of cosmological parameters and 
marginalize over $b$ as a nuisance parameter. The Fisher constraints 
for the near future are presented in the left panel of Fig.~\ref{BiasFig},
while the right  panel shows the constraints for the far future. 
In both cases, the red open circles show the constraints of Voids + 
PLANCK + HST from the upper panel of Fig.~\ref{Constraints}, while the 
solid thin black line shows the constraints from Voids +PLANCK + HST +SNE, 
where it was assumed that $b=1$. The green open squares show the 
corresponding constraints for Voids + HST + PLANCK, and the thick blue 
solid ellipses show the constraints for Voids + HST +PLANCK + SNE, 
when the bias is marginalized over. \\ 

\begin{figure*}[t]
\begin{center}
\includegraphics[width=0.95\textwidth]{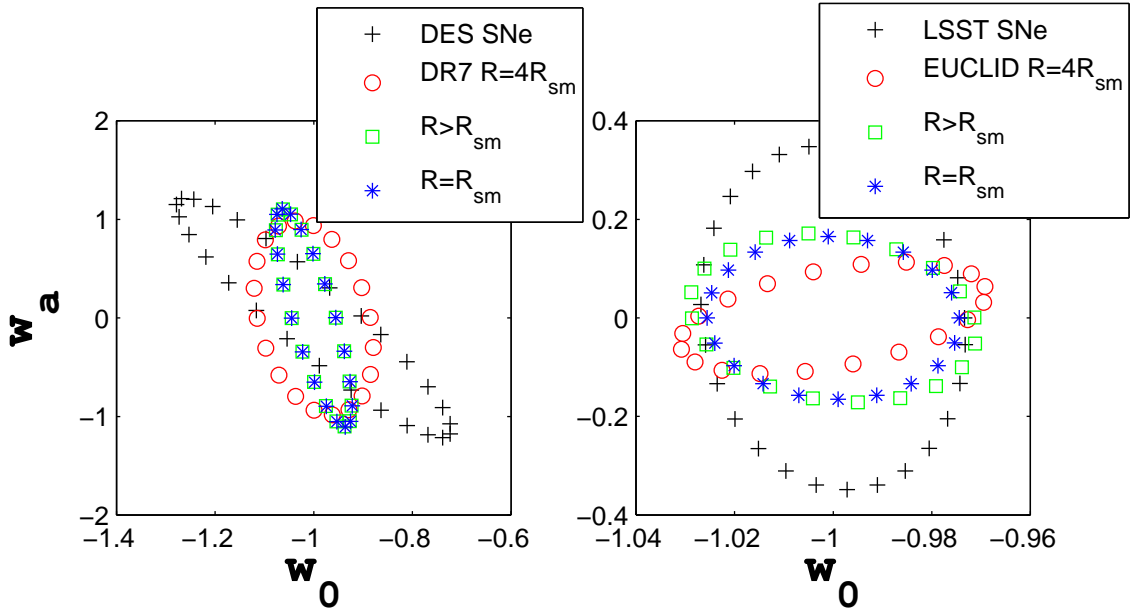}
\caption{{\textbf{Sensitivity of Fisher Constraints with respect to the prescription
of Void Selection:}} Comparison of the constraints (red open circles) from 
voids shown in Fig.~\ref{Constraints} with other prescriptions. PLANCK and HST priors were used in all these plots. The other
prescriptions lead to better constraints} 
\label{Degradation_Algorithm}
\end{center}
\end{figure*}

{(c) \it{Void Selection Prescription}} 
While the eigenvalues of the void ellipsoid are expected to trace the 
eigenvalues of the tidal ellipsoid, the eigenvalues themselves are 
stochastic quantities and the connection to theory comes from studying the
distribution of these eigenvalues. Hence it is important to select a set of
voids from the data that will accurately reflect the theoretical 
distribution computed. 
As discussed in ~\citet{2008MNRAS.387..933C}, the void finders available 
use different methods to identify voids, and these result in different
definitions of voids. A number of these void finders are based on 
demarcating contiguous regions of space of different shapes through some 
variant of a clustering algorithm, while other void finders like 
~\citet{2009arXiv0906.4101L}
identify voids from a density field smoothed at a particular length scale. 
On the theoretical side, we can compute the probability distribution of 
the eigenvalues of the tidal tensor analytically through the Doroshkevich
formula Eqn.~\ref{DKV}, which we use in the computations here,
which is the distribution valid at all points in space rather than at voids
 in particular. One may also compute the distribution of the eigenvalues
(i) for a void of size $R$ identified with the size of the fluctuation at 
shell crossing as shown in subsections ~\ref{distribution_eigenvalues} and 
~\ref{evolution_eigenvalues}, or (ii)
 at the minima of the density field when smoothed at a particular length 
scale (eg. see Appendix B of ~\citet{2009arXiv0906.4101L}).
Both of these are not analytic estimates, but they can used to construct 
samples of the eigenvalue distributions using Monte Carlo methods and lend 
themselves naturally to use with the two classes of void finders 
respectively. The use of computationally intensive Monte Carlo is beyond 
the scope of this paper based on Fisher estimates. Instead we use the 
analytic Doroshkevich formula which was shown to be close to both of these
distributions, but this requires us to identify the set of voids that 
correspond to the voids obtained by smoothing the density field at a 
particular Lagrangian scale $R_{\text{Smooth}}$. 
If we find a set of voids at a particular redshift of a set of different
sizes, how can we identify what smoothing scale these voids correspond to?
Given a set of point particles in space, we understand 
the action of smoothing: it tends to homogenize the field at scales below 
the smoothing scale. 
Thus, one may expect that on smoothing by a scale $R_{\text{Smooth}}$, one will 
be left with voids with distribution such that there are few voids of size
 below $\approx R_{\text{Smooth}},$ while the smoothing operation may slightly 
modify the shapes and sizes voids of larger size. 
At a particular redshift, the probability of forming
large voids is much smaller than forming smaller voids. 
Consequently, the distribution of sizes of voids when the density field 
is smoothed to a scale $R_{\text{Smooth}}$, should be peaked at $\sim R_{\text{Smooth}}.$
From simulations used in ~\citet{2009arXiv0906.4101L}, it appears that the 
distribution of the number of voids with radius $R$
in a density field smoothed by a filter of size $R_{	\text{Smooth}}$, is peaked 
at $R \approx 4 R_{\text{Smooth}}$ and falls off rapidly above that.
While this inspired our choice for identification of voids, it is important
 to keep in mind that the distribution depends on the cosmological 
parameters through $\sigma_{lin}(R,z)$. Consequently using an inaccurate 
selection 
criterion for voids can introduce biases in parameter estimation, and the
correct prescription may also change the errors and constraints. 
In order to get a sense for how severely the 
constraints might be degraded when this is done, we compute the constraints
for three different prescriptions of identification the set of voids and
compare how far the constraints are degraded in different cases that 
suggest themselves. From the right panel of 
Fig.~\ref{GrowthDerivs}, we see that the distribution gets broader for 
 larger values of $\sigma_R$. Since this corresponds to lower theoretical 
predictability, we should expect the parameter  
constraints to get degraded as the filtering scale $R$ becomes smaller. On the other 
hand, this will lead to a larger number of voids since there are many more
smaller voids.\\

One may expect that when the density field is smoothed at $R_{\text{Smooth}}$,
a non-negligible fraction of the voids have radii between $R_{\text{Smooth}}$ 
and $4R_{\text{Smooth}}$. We can therefore use a different limit $R = R_{\text{Smooth}}$
in accordance with our calculations using the generalized excursion set
formalism in subsection. ~\ref{distribution_eigenvalues}.
Finally, if we assume that all voids larger than a particular smoothing 
scale would be found, we can take $R_{\text{Smooth}}= Min(\{R\})$ 
found in that redshift bin. This is similar to the method adopted
by ~\citet{2007arXiv0704.0881L}.
The corresponding constraints are shown in 
Fig.~\ref{Degradation_Algorithm}. The red open circles show the constraints
shown in Fig.~\ref{Constraints} for the prescription where 
$R_{	\text{Smooth}}= R/4,$ while the blue asterisks show the constraints 
obtained for the case where $R_{\text{Smooth}} = R$, and the open green squares
show the constraints for the case where $R_{\text{Smooth}}=Min(\{R\})$. \\

\begin{figure*}
\begin{center}
\includegraphics[width=0.95\textwidth]{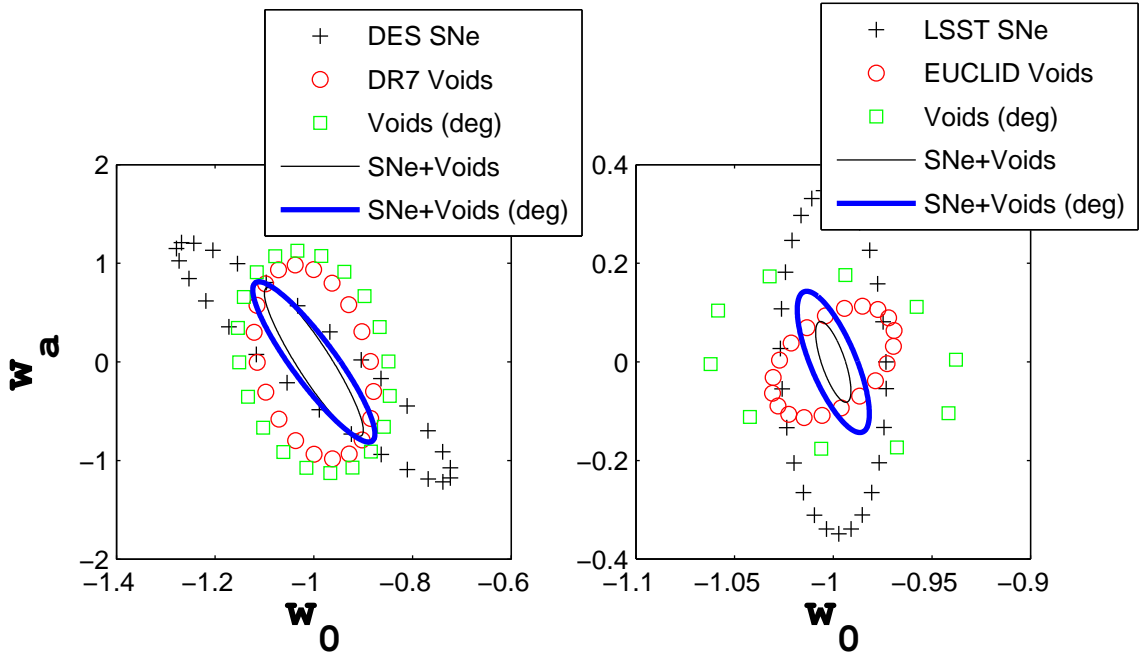}
\caption{\textbf{Sensitivity of Fisher Constraints with $\sigma_\epsilon$:}
Comparison of constraints from voids (open red circles) 
with $\sigma_{\epsilon}=0.1$ shown in Fig.~\ref{Constraints} with 
pessimistically degraded constraints from voids due to a larger $\sigma_{\epsilon}=0.4$ for
 data in the near future (left panel) and in the far future (right panel).
Despite the degradation, the constraints are still interesting. 
HST and PLANCK priors were used in all the plots.
}
\label{VariationWithSigmaEpsilon}
\end{center}
\end{figure*}
{(d)\it{Sensitivity to Error Levels}}

As discussed before, in our method of forecasting for 
Fig.~\ref{Constraints}, we have used a Gaussian  Likelihood with an error 
$\sigma_{\epsilon}=0.1$ assuming that its order was set by the 
uncertainty of measuring the void size which was limited by the size of the
void shell (if $\Delta \approx 0.4$). Indeed, this seems larger than the 
values of the error levels computed in section 5.3.2 of ~\citet{2009arXiv0906.4101L}.
Further, in our analysis, we have 
assumed that the ellipticities of the mass tensor
of voids are perfect tracers of ellipticity of the tidal tensor. 
More realistically, there would be some scatter around the correlation as
shown in section 5.2 of ~\citet{2009arXiv0906.4101L}. It is quite possible that scatter
of this kind, or the assumptions that we have made might increase the
level of error bars on $\epsilon$ quantitatively. Therefore, 
we investigate the sensitivity of the constraints to the value of 
$\sigma_{\epsilon},$ the error to which the ellipticity was assumed to be 
measured.\\ 
We show these constraints in Fig.~\ref{VariationWithSigmaEpsilon}, where 
the contours with red open circles show the constraints using Voids +
PLANCK + HST shown in the upper panels of Fig.~\ref{Constraints} with 
$A=1$ and $\sigma_{\epsilon}=0.1$, while the open green squares are
the constraints where $\sigma_{epsilon}$ has been increased to $0.4$. 
The solid lines show the constraints where the constraints are estimated 
with simultaneous use of the SNe data, ie. DES SNE for the left panel and 
LSST SNe for the right panel. The thin black solid line is for 
$\sigma_{\epsilon}=0.1$, while the thick blue solid line is for 
$\sigma_{\epsilon}=0.4$. The 
contours in black "+" symbols show the constraints from SNe + PLANCK + HST 
for reference.  

\begin{table*}[!h]
\caption{Relative Figure of Merit (FoM) for using voids}
\label{table:FoM}
\begin{tabular}{|c|c|c|c|c|}
\hline
\multicolumn{1}{|c|}{}&\multicolumn{2}{|c|}{SDSS+DES+HST+PLANCK} &\multicolumn{2}{|c|}{EUCLID+LSST+HST+PLANCK}\\
\hline
Parameters & Voids+CMB+HST & 
Voids + CMB+ HST +SNE& Voids+CMB+HST& Voids+CMB+HST+SNE\\
\hline
$A=1,\sigma=0.1$ & 1.2 &  16.8 &  8.8&  331.0 \\
\hline
$A=2,\sigma =0.1$ & 0.6 & 13.3 & 0.5 & 21.3 \\
\hline
$A =1,\sigma=0.4$ & 0.5 & 7.7 & 0.7 & 27.6 \\
\hline
Marginalized over $b$ & 0.2 &  3.3 & 0.2 & 104.5 \\
\hline
$R_{\text{Smooth}}= Min(\{R\})$ & 6.1 & 24.9 & 3.6 & 73.0 \\
\hline
$R_{\text{Smooth}}= R$ & 6.1 & 24.7 & 4.8 & 85.2 \\
\hline
\end{tabular}
\end{table*}
\section{Summary and Discussions}
The growth of cosmic structures with time depends on the background 
cosmology. Consequently, the growth of structures have been used to 
constrain the parameters of the background cosmology. Traditionally,
the measures of growth used have characterized the growth of the 
volume of fluctuations. However, since the fluctuations are not 
individually isotropic, there is further information about the cosmology
in the growth of asymmetry of the structures which could be extracted from
its shape. Such a quantity parametrizing the shape of voids and its
evolution was studied in ~\citet{2007PhRvL..98h1301P,2007arXiv0704.0881L}. 
The basic idea is that 
void shapes can be approximated as ellipsoidal structures, and 
relative sizes of the principal axes can be used as tracers of functions
of eigenvalues of the tidal ellipsoid.
In a spectroscopic survey, all three axes of the void ellipsoid may be 
measured, and thus asymmetry parameters which describe the shape of the 
ellipsoid are related to the quantities involving the eigenvalues of the 
tidal tensor, which depend on the background cosmology through the linearly
extrapolated variance in fluctuations. Such spectroscopic surveys have been
planned for studying large scale structure using traditional methods; thus
 the use of shapes does not necessarily require new surveys, but allows one
to leverage data in an additional way. 
{\citet{2009arXiv0906.4101L} show that 
recovering the the tidal ellipticity of voids to high precision is indeed 
feasible. To
 do so, they identify voids and characterize the void tidal ellipticity 
using the simulated galaxy positions derived from a numerical simulation.
 These derived 
ellipticities are then compared to the tidal ellipticity of the complete displacement field given by the simulation.

In this paper, we study the constraints on dark energy parameters from 
future surveys in terms of Fisher forecasts. The 
likelihood is a strong function of the linearly extrapolated 
variance of fluctuations at the redshift of the void at the scale of the 
Lagrangian size of the void. Since voids expand in comoving coordinates, 
their Lagrangian size is smaller than their observed
(comoving) size, and this corresponds to a larger variance.
variance at a smaller scale than the observed void size. 
We assume an error model with Gaussian noise on the measured ellipticity of
the voids, and an arbitrarily assumed error on the ellipticity. We 
provide explicit formulae for Fisher matrices, and an estimate of 
the number of voids  expected to be found from planned future surveys using
semi-analytic methods. By 
comparing these Fisher constraints using void shapes from these surveys to
 the traditional constraints from other measures, we find this method to 
be promising: the constraints are quite competitive with traditional probes
in the near future and combining the 
constraints with supernovae data improves the DETF  Figure of Merit for the
supernovae data by a factor of about ten. For futuristic data, we find that
the constraints are close to ten times better than supernovae data,
and combining with supernovae data, we can improve the FoM by a factor of 
a few hundred.  

We have used the Doroshkevich formula for the ellipticity 
throughout, but it has been shown~\citep{2009arXiv0906.4101L} that 
the distribution of ellipticity for a minima in the density field is 
slightly different. In actual parameter estimation, we will have to
account for this. We shall also have to use the scatter in the correlation
of the ellipticity of the void ellipsoid with the real shape of the tidal
tensor as obtained from specific void identification algorithms. An issue
we have not addressed here is the ellipticity of voids that can be 
generated due to redshift distortions 
~\citep{1995ApJ...452...25R,1996ApJ...470..160R} which would have to be 
modeled to obtain unbiased parameter constraints from voids.

The Fisher constraints are computed using simple models of dynamics and 
a likelihood. For estimation of parameters, each of these would need to 
be computed precisely. In the subsection ~\ref{ResultsIssues}, we discuss 
some of the main sources of errors and ambiguities in our forecasts. 
We indicate how more rigorous, though computationally intensive
calculations may be devised. We attempt to estimate how the parameter
constraints might be affected by these more rigorous methods. While the
constraints are often weakened, they still remain at least competitive 
with other constraints in the near future and the far future. In the case 
of futuristic surveys, addition of the void ellipticity to other 
constraints result in an improvement of the FoM by a factor of at least a 
hundred, in spite of degradation due to additional systematics. 
We therefore feel that our study makes a strong case for pursuing this idea in 
greater detail.

\section{Acknowledgments}
We would like to thank G.~Lavaux for many useful discussions and 
sharing insights from his results that motivated some choices in this 
paper. 
RB would like to thank W.M.~Wood-Vasey for discussions about LSST 
supernovae. 
The authors would
like to thank the California Institute of Technology for hospitality during
which part of this work was done. The authors acknowledge financial 
support from NSF grantAST 07-08849.
\bibliography{VoidPaper.bbl}

\begin{thebibliography}{79}
\expandafter\ifx\csname natexlab\endcsname\relax\def\natexlab#1{#1}\fi
\expandafter\ifx\csname bibnamefont\endcsname\relax
  \def\bibnamefont#1{#1}\fi
\expandafter\ifx\csname bibfnamefont\endcsname\relax
  \def\bibfnamefont#1{#1}\fi
\expandafter\ifx\csname citenamefont\endcsname\relax
  \def\citenamefont#1{#1}\fi
\expandafter\ifx\csname url\endcsname\relax
  \def\url#1{\texttt{#1}}\fi
\expandafter\ifx\csname urlprefix\endcsname\relax\def\urlprefix{URL }\fi
\providecommand{\bibinfo}[2]{#2}
\providecommand{\eprint}[2][]{\url{#2}}

\bibitem[{\citenamefont{{Perlmutter} et~al.}(1999)\citenamefont{{Perlmutter},
  {Aldering}, {Goldhaber}, {Knop}, {Nugent}, {Castro}, {Deustua}, {Fabbro},
  {Goobar}, {Groom} et~al.}}]{1999ApJ...517..565P}
\bibinfo{author}{\bibfnamefont{S.}~\bibnamefont{{Perlmutter}}},
  \bibinfo{author}{\bibfnamefont{G.}~\bibnamefont{{Aldering}}},
  \bibinfo{author}{\bibfnamefont{G.}~\bibnamefont{{Goldhaber}}},
  \bibinfo{author}{\bibfnamefont{R.~A.} \bibnamefont{{Knop}}},
  \bibinfo{author}{\bibfnamefont{P.}~\bibnamefont{{Nugent}}},
  \bibinfo{author}{\bibfnamefont{P.~G.} \bibnamefont{{Castro}}},
  \bibinfo{author}{\bibfnamefont{S.}~\bibnamefont{{Deustua}}},
  \bibinfo{author}{\bibfnamefont{S.}~\bibnamefont{{Fabbro}}},
  \bibinfo{author}{\bibfnamefont{A.}~\bibnamefont{{Goobar}}},
  \bibinfo{author}{\bibfnamefont{D.~E.} \bibnamefont{{Groom}}},
  \bibnamefont{et~al.}, \bibinfo{journal}{\apj} \textbf{\bibinfo{volume}{517}},
  \bibinfo{pages}{565} (\bibinfo{year}{1999}), \eprint{arXiv:astro-ph/9812133}.

\bibitem[{\citenamefont{{Riess} et~al.}(1998)\citenamefont{{Riess},
  {Filippenko}, {Challis}, {Clocchiatti}, {Diercks}, {Garnavich}, {Gilliland},
  {Hogan}, {Jha}, {Kirshner} et~al.}}]{1998AJ....116.1009R}
\bibinfo{author}{\bibfnamefont{A.~G.} \bibnamefont{{Riess}}},
  \bibinfo{author}{\bibfnamefont{A.~V.} \bibnamefont{{Filippenko}}},
  \bibinfo{author}{\bibfnamefont{P.}~\bibnamefont{{Challis}}},
  \bibinfo{author}{\bibfnamefont{A.}~\bibnamefont{{Clocchiatti}}},
  \bibinfo{author}{\bibfnamefont{A.}~\bibnamefont{{Diercks}}},
  \bibinfo{author}{\bibfnamefont{P.~M.} \bibnamefont{{Garnavich}}},
  \bibinfo{author}{\bibfnamefont{R.~L.} \bibnamefont{{Gilliland}}},
  \bibinfo{author}{\bibfnamefont{C.~J.} \bibnamefont{{Hogan}}},
  \bibinfo{author}{\bibfnamefont{S.}~\bibnamefont{{Jha}}},
  \bibinfo{author}{\bibfnamefont{R.~P.} \bibnamefont{{Kirshner}}},
  \bibnamefont{et~al.}, \bibinfo{journal}{\aj} \textbf{\bibinfo{volume}{116}},
  \bibinfo{pages}{1009} (\bibinfo{year}{1998}),
  \eprint{arXiv:astro-ph/9805201}.

\bibitem[{\citenamefont{{Garnavich} et~al.}(1998)\citenamefont{{Garnavich},
  {Jha}, {Challis}, {Clocchiatti}, {Diercks}, {Filippenko}, {Gilliland},
  {Hogan}, {Kirshner}, {Leibundgut} et~al.}}]{1998ApJ...509...74G}
\bibinfo{author}{\bibfnamefont{P.~M.} \bibnamefont{{Garnavich}}},
  \bibinfo{author}{\bibfnamefont{S.}~\bibnamefont{{Jha}}},
  \bibinfo{author}{\bibfnamefont{P.}~\bibnamefont{{Challis}}},
  \bibinfo{author}{\bibfnamefont{A.}~\bibnamefont{{Clocchiatti}}},
  \bibinfo{author}{\bibfnamefont{A.}~\bibnamefont{{Diercks}}},
  \bibinfo{author}{\bibfnamefont{A.~V.} \bibnamefont{{Filippenko}}},
  \bibinfo{author}{\bibfnamefont{R.~L.} \bibnamefont{{Gilliland}}},
  \bibinfo{author}{\bibfnamefont{C.~J.} \bibnamefont{{Hogan}}},
  \bibinfo{author}{\bibfnamefont{R.~P.} \bibnamefont{{Kirshner}}},
  \bibinfo{author}{\bibfnamefont{B.}~\bibnamefont{{Leibundgut}}},
  \bibnamefont{et~al.}, \bibinfo{journal}{\apj} \textbf{\bibinfo{volume}{509}},
  \bibinfo{pages}{74} (\bibinfo{year}{1998}), \eprint{arXiv:astro-ph/9806396}.

\bibitem[{\citenamefont{{Knop} et~al.}(2003)\citenamefont{{Knop}, {Aldering},
  {Amanullah}, {Astier}, {Blanc}, {Burns}, {Conley}, {Deustua}, {Doi}, {Ellis}
  et~al.}}]{2003ApJ...598..102K}
\bibinfo{author}{\bibfnamefont{R.~A.} \bibnamefont{{Knop}}},
  \bibinfo{author}{\bibfnamefont{G.}~\bibnamefont{{Aldering}}},
  \bibinfo{author}{\bibfnamefont{R.}~\bibnamefont{{Amanullah}}},
  \bibinfo{author}{\bibfnamefont{P.}~\bibnamefont{{Astier}}},
  \bibinfo{author}{\bibfnamefont{G.}~\bibnamefont{{Blanc}}},
  \bibinfo{author}{\bibfnamefont{M.~S.} \bibnamefont{{Burns}}},
  \bibinfo{author}{\bibfnamefont{A.}~\bibnamefont{{Conley}}},
  \bibinfo{author}{\bibfnamefont{S.~E.} \bibnamefont{{Deustua}}},
  \bibinfo{author}{\bibfnamefont{M.}~\bibnamefont{{Doi}}},
  \bibinfo{author}{\bibfnamefont{R.}~\bibnamefont{{Ellis}}},
  \bibnamefont{et~al.}, \bibinfo{journal}{\apj} \textbf{\bibinfo{volume}{598}},
  \bibinfo{pages}{102} (\bibinfo{year}{2003}), \eprint{arXiv:astro-ph/0309368}.

\bibitem[{\citenamefont{{Tonry} et~al.}(2003)\citenamefont{{Tonry}, {Schmidt},
  {Barris}, {Candia}, {Challis}, {Clocchiatti}, {Coil}, {Filippenko},
  {Garnavich}, {Hogan} et~al.}}]{2003ApJ...594....1T}
\bibinfo{author}{\bibfnamefont{J.~L.} \bibnamefont{{Tonry}}},
  \bibinfo{author}{\bibfnamefont{B.~P.} \bibnamefont{{Schmidt}}},
  \bibinfo{author}{\bibfnamefont{B.}~\bibnamefont{{Barris}}},
  \bibinfo{author}{\bibfnamefont{P.}~\bibnamefont{{Candia}}},
  \bibinfo{author}{\bibfnamefont{P.}~\bibnamefont{{Challis}}},
  \bibinfo{author}{\bibfnamefont{A.}~\bibnamefont{{Clocchiatti}}},
  \bibinfo{author}{\bibfnamefont{A.~L.} \bibnamefont{{Coil}}},
  \bibinfo{author}{\bibfnamefont{A.~V.} \bibnamefont{{Filippenko}}},
  \bibinfo{author}{\bibfnamefont{P.}~\bibnamefont{{Garnavich}}},
  \bibinfo{author}{\bibfnamefont{C.}~\bibnamefont{{Hogan}}},
  \bibnamefont{et~al.}, \bibinfo{journal}{\apj} \textbf{\bibinfo{volume}{594}},
  \bibinfo{pages}{1} (\bibinfo{year}{2003}), \eprint{arXiv:astro-ph/0305008}.

\bibitem[{\citenamefont{{Riess} et~al.}(2004)\citenamefont{{Riess}, {Strolger},
  {Tonry}, {Casertano}, {Ferguson}, {Mobasher}, {Challis}, {Filippenko}, {Jha},
  {Li} et~al.}}]{2004ApJ...607..665R}
\bibinfo{author}{\bibfnamefont{A.~G.} \bibnamefont{{Riess}}},
  \bibinfo{author}{\bibfnamefont{L.-G.} \bibnamefont{{Strolger}}},
  \bibinfo{author}{\bibfnamefont{J.}~\bibnamefont{{Tonry}}},
  \bibinfo{author}{\bibfnamefont{S.}~\bibnamefont{{Casertano}}},
  \bibinfo{author}{\bibfnamefont{H.~C.} \bibnamefont{{Ferguson}}},
  \bibinfo{author}{\bibfnamefont{B.}~\bibnamefont{{Mobasher}}},
  \bibinfo{author}{\bibfnamefont{P.}~\bibnamefont{{Challis}}},
  \bibinfo{author}{\bibfnamefont{A.~V.} \bibnamefont{{Filippenko}}},
  \bibinfo{author}{\bibfnamefont{S.}~\bibnamefont{{Jha}}},
  \bibinfo{author}{\bibfnamefont{W.}~\bibnamefont{{Li}}}, \bibnamefont{et~al.},
  \bibinfo{journal}{\apj} \textbf{\bibinfo{volume}{607}}, \bibinfo{pages}{665}
  (\bibinfo{year}{2004}), \eprint{arXiv:astro-ph/0402512}.

\bibitem[{\citenamefont{{Astier} et~al.}(2006)\citenamefont{{Astier}, {Guy},
  {Regnault}, {Pain}, {Aubourg}, {Balam}, {Basa}, {Carlberg}, {Fabbro},
  {Fouchez} et~al.}}]{2006A&A...447...31A}
\bibinfo{author}{\bibfnamefont{P.}~\bibnamefont{{Astier}}},
  \bibinfo{author}{\bibfnamefont{J.}~\bibnamefont{{Guy}}},
  \bibinfo{author}{\bibfnamefont{N.}~\bibnamefont{{Regnault}}},
  \bibinfo{author}{\bibfnamefont{R.}~\bibnamefont{{Pain}}},
  \bibinfo{author}{\bibfnamefont{E.}~\bibnamefont{{Aubourg}}},
  \bibinfo{author}{\bibfnamefont{D.}~\bibnamefont{{Balam}}},
  \bibinfo{author}{\bibfnamefont{S.}~\bibnamefont{{Basa}}},
  \bibinfo{author}{\bibfnamefont{R.~G.} \bibnamefont{{Carlberg}}},
  \bibinfo{author}{\bibfnamefont{S.}~\bibnamefont{{Fabbro}}},
  \bibinfo{author}{\bibfnamefont{D.}~\bibnamefont{{Fouchez}}},
  \bibnamefont{et~al.}, \bibinfo{journal}{\aap} \textbf{\bibinfo{volume}{447}},
  \bibinfo{pages}{31} (\bibinfo{year}{2006}), \eprint{arXiv:astro-ph/0510447}.

\bibitem[{\citenamefont{{Wood-Vasey} et~al.}(2007)\citenamefont{{Wood-Vasey},
  {Miknaitis}, {Stubbs}, {Jha}, {Riess}, {Garnavich}, {Kirshner}, {Aguilera},
  {Becker}, {Blackman} et~al.}}]{2007ApJ...666..694W}
\bibinfo{author}{\bibfnamefont{W.~M.} \bibnamefont{{Wood-Vasey}}},
  \bibinfo{author}{\bibfnamefont{G.}~\bibnamefont{{Miknaitis}}},
  \bibinfo{author}{\bibfnamefont{C.~W.} \bibnamefont{{Stubbs}}},
  \bibinfo{author}{\bibfnamefont{S.}~\bibnamefont{{Jha}}},
  \bibinfo{author}{\bibfnamefont{A.~G.} \bibnamefont{{Riess}}},
  \bibinfo{author}{\bibfnamefont{P.~M.} \bibnamefont{{Garnavich}}},
  \bibinfo{author}{\bibfnamefont{R.~P.} \bibnamefont{{Kirshner}}},
  \bibinfo{author}{\bibfnamefont{C.}~\bibnamefont{{Aguilera}}},
  \bibinfo{author}{\bibfnamefont{A.~C.} \bibnamefont{{Becker}}},
  \bibinfo{author}{\bibfnamefont{J.~W.} \bibnamefont{{Blackman}}},
  \bibnamefont{et~al.}, \bibinfo{journal}{\apj} \textbf{\bibinfo{volume}{666}},
  \bibinfo{pages}{694} (\bibinfo{year}{2007}), \eprint{arXiv:astro-ph/0701041}.

\bibitem[{\citenamefont{{Hicken} et~al.}(2009)\citenamefont{{Hicken},
  {Wood-Vasey}, {Blondin}, {Challis}, {Jha}, {Kelly}, {Rest}, and
  {Kirshner}}}]{2009arXiv0901.4804H}
\bibinfo{author}{\bibfnamefont{M.}~\bibnamefont{{Hicken}}},
  \bibinfo{author}{\bibfnamefont{W.~M.} \bibnamefont{{Wood-Vasey}}},
  \bibinfo{author}{\bibfnamefont{S.}~\bibnamefont{{Blondin}}},
  \bibinfo{author}{\bibfnamefont{P.}~\bibnamefont{{Challis}}},
  \bibinfo{author}{\bibfnamefont{S.}~\bibnamefont{{Jha}}},
  \bibinfo{author}{\bibfnamefont{P.~L.} \bibnamefont{{Kelly}}},
  \bibinfo{author}{\bibfnamefont{A.}~\bibnamefont{{Rest}}}, \bibnamefont{and}
  \bibinfo{author}{\bibfnamefont{R.~P.} \bibnamefont{{Kirshner}}},
  \bibinfo{journal}{ArXiv e-prints}  (\bibinfo{year}{2009}),
  \eprint{0901.4804}.

\bibitem[{\citenamefont{{Weinberg}}(1989)}]{1989RvMP...61....1W}
\bibinfo{author}{\bibfnamefont{S.}~\bibnamefont{{Weinberg}}},
  \bibinfo{journal}{Reviews of Modern Physics} \textbf{\bibinfo{volume}{61}},
  \bibinfo{pages}{1} (\bibinfo{year}{1989}).

\bibitem[{\citenamefont{{Carroll} et~al.}(1992)\citenamefont{{Carroll},
  {Press}, and {Turner}}}]{1992ARA&A..30..499C}
\bibinfo{author}{\bibfnamefont{S.~M.} \bibnamefont{{Carroll}}},
  \bibinfo{author}{\bibfnamefont{W.~H.} \bibnamefont{{Press}}},
  \bibnamefont{and} \bibinfo{author}{\bibfnamefont{E.~L.}
  \bibnamefont{{Turner}}}, \bibinfo{journal}{\araa}
  \textbf{\bibinfo{volume}{30}}, \bibinfo{pages}{499} (\bibinfo{year}{1992}).

\bibitem[{\citenamefont{{Weinberg}}(2000)}]{2000astro.ph..5265W}
\bibinfo{author}{\bibfnamefont{S.}~\bibnamefont{{Weinberg}}},
  \bibinfo{journal}{ArXiv Astrophysics e-prints}  (\bibinfo{year}{2000}),
  \eprint{arXiv:astro-ph/0005265}.

\bibitem[{\citenamefont{{Carroll}}(2001)}]{2001LRR.....4....1C}
\bibinfo{author}{\bibfnamefont{S.~M.} \bibnamefont{{Carroll}}},
  \bibinfo{journal}{Living Reviews in Relativity} \textbf{\bibinfo{volume}{4}},
  \bibinfo{pages}{1} (\bibinfo{year}{2001}), \eprint{arXiv:astro-ph/0004075}.

\bibitem[{\citenamefont{{Carroll}}(2004)}]{2004mmu..symp..235C}
\bibinfo{author}{\bibfnamefont{S.~M.} \bibnamefont{{Carroll}}}, pp.
  \bibinfo{pages}{235--+} (\bibinfo{year}{2004}).

\bibitem[{\citenamefont{{Buchert} et~al.}(2000)\citenamefont{{Buchert},
  {Kerscher}, and {Sicka}}}]{2000PhRvD..62d3525B}
\bibinfo{author}{\bibfnamefont{T.}~\bibnamefont{{Buchert}}},
  \bibinfo{author}{\bibfnamefont{M.}~\bibnamefont{{Kerscher}}},
  \bibnamefont{and} \bibinfo{author}{\bibfnamefont{C.}~\bibnamefont{{Sicka}}},
  \bibinfo{journal}{\prd} \textbf{\bibinfo{volume}{62}},
  \bibinfo{pages}{043525} (\bibinfo{year}{2000}),
  \eprint{arXiv:astro-ph/9912347}.

\bibitem[{\citenamefont{{Kolb} et~al.}(2005)\citenamefont{{Kolb}, {Matarrese},
  {Notari}, and {Riotto}}}]{2005PhRvD..71b3524K}
\bibinfo{author}{\bibfnamefont{E.~W.} \bibnamefont{{Kolb}}},
  \bibinfo{author}{\bibfnamefont{S.}~\bibnamefont{{Matarrese}}},
  \bibinfo{author}{\bibfnamefont{A.}~\bibnamefont{{Notari}}}, \bibnamefont{and}
  \bibinfo{author}{\bibfnamefont{A.}~\bibnamefont{{Riotto}}},
  \bibinfo{journal}{\prd} \textbf{\bibinfo{volume}{71}},
  \bibinfo{pages}{023524} (\bibinfo{year}{2005}),
  \eprint{arXiv:hep-ph/0409038}.

\bibitem[{\citenamefont{{Ellis} and {Buchert}}(2005)}]{2005PhLA..347...38E}
\bibinfo{author}{\bibfnamefont{G.~F.~R.} \bibnamefont{{Ellis}}}
  \bibnamefont{and}
  \bibinfo{author}{\bibfnamefont{T.}~\bibnamefont{{Buchert}}},
  \bibinfo{journal}{Physics Letters A} \textbf{\bibinfo{volume}{347}},
  \bibinfo{pages}{38} (\bibinfo{year}{2005}), \eprint{arXiv:gr-qc/0506106}.

\bibitem[{\citenamefont{{Kolb} et~al.}(2006)\citenamefont{{Kolb}, {Matarrese},
  and {Riotto}}}]{2006NJPh....8..322K}
\bibinfo{author}{\bibfnamefont{E.~W.} \bibnamefont{{Kolb}}},
  \bibinfo{author}{\bibfnamefont{S.}~\bibnamefont{{Matarrese}}},
  \bibnamefont{and} \bibinfo{author}{\bibfnamefont{A.}~\bibnamefont{{Riotto}}},
  \bibinfo{journal}{New Journal of Physics} \textbf{\bibinfo{volume}{8}},
  \bibinfo{pages}{322} (\bibinfo{year}{2006}), \eprint{arXiv:astro-ph/0506534}.

\bibitem[{\citenamefont{{Dvali} et~al.}(2000)\citenamefont{{Dvali},
  {Gabadadze}, and {Porrati}}}]{2000PhLB..485..208D}
\bibinfo{author}{\bibfnamefont{G.}~\bibnamefont{{Dvali}}},
  \bibinfo{author}{\bibfnamefont{G.}~\bibnamefont{{Gabadadze}}},
  \bibnamefont{and}
  \bibinfo{author}{\bibfnamefont{M.}~\bibnamefont{{Porrati}}},
  \bibinfo{journal}{Physics Letters B} \textbf{\bibinfo{volume}{485}},
  \bibinfo{pages}{208} (\bibinfo{year}{2000}), \eprint{arXiv:hep-th/0005016}.

\bibitem[{\citenamefont{{Carroll} et~al.}(2005)\citenamefont{{Carroll}, {de
  Felice}, {Duvvuri}, {Easson}, {Trodden}, and {Turner}}}]{2005PhRvD..71f3513C}
\bibinfo{author}{\bibfnamefont{S.~M.} \bibnamefont{{Carroll}}},
  \bibinfo{author}{\bibfnamefont{A.}~\bibnamefont{{de Felice}}},
  \bibinfo{author}{\bibfnamefont{V.}~\bibnamefont{{Duvvuri}}},
  \bibinfo{author}{\bibfnamefont{D.~A.} \bibnamefont{{Easson}}},
  \bibinfo{author}{\bibfnamefont{M.}~\bibnamefont{{Trodden}}},
  \bibnamefont{and} \bibinfo{author}{\bibfnamefont{M.~S.}
  \bibnamefont{{Turner}}}, \bibinfo{journal}{\prd}
  \textbf{\bibinfo{volume}{71}}, \bibinfo{pages}{063513}
  (\bibinfo{year}{2005}), \eprint{arXiv:astro-ph/0410031}.

\bibitem[{\citenamefont{{Jain} and {Zhang}}(2008)}]{2008PhRvD..78f3503J}
\bibinfo{author}{\bibfnamefont{B.}~\bibnamefont{{Jain}}} \bibnamefont{and}
  \bibinfo{author}{\bibfnamefont{P.}~\bibnamefont{{Zhang}}},
  \bibinfo{journal}{\prd} \textbf{\bibinfo{volume}{78}},
  \bibinfo{pages}{063503} (\bibinfo{year}{2008}), \eprint{0709.2375}.

\bibitem[{\citenamefont{{Freedman} et~al.}(2001)\citenamefont{{Freedman},
  {Madore}, {Gibson}, {Ferrarese}, {Kelson}, {Sakai}, {Mould}, {Kennicutt},
  {Ford}, {Graham} et~al.}}]{2001ApJ...553...47F}
\bibinfo{author}{\bibfnamefont{W.~L.} \bibnamefont{{Freedman}}},
  \bibinfo{author}{\bibfnamefont{B.~F.} \bibnamefont{{Madore}}},
  \bibinfo{author}{\bibfnamefont{B.~K.} \bibnamefont{{Gibson}}},
  \bibinfo{author}{\bibfnamefont{L.}~\bibnamefont{{Ferrarese}}},
  \bibinfo{author}{\bibfnamefont{D.~D.} \bibnamefont{{Kelson}}},
  \bibinfo{author}{\bibfnamefont{S.}~\bibnamefont{{Sakai}}},
  \bibinfo{author}{\bibfnamefont{J.~R.} \bibnamefont{{Mould}}},
  \bibinfo{author}{\bibfnamefont{R.~C.} \bibnamefont{{Kennicutt}},
  \bibfnamefont{Jr.}}, \bibinfo{author}{\bibfnamefont{H.~C.}
  \bibnamefont{{Ford}}}, \bibinfo{author}{\bibfnamefont{J.~A.}
  \bibnamefont{{Graham}}}, \bibnamefont{et~al.}, \bibinfo{journal}{\apj}
  \textbf{\bibinfo{volume}{553}}, \bibinfo{pages}{47} (\bibinfo{year}{2001}),
  \eprint{arXiv:astro-ph/0012376}.

\bibitem[{\citenamefont{{Cole} et~al.}(2005)\citenamefont{{Cole}, {Percival},
  {Peacock}, {Norberg}, {Baugh}, {Frenk}, {Baldry}, {Bland-Hawthorn},
  {Bridges}, {Cannon} et~al.}}]{2005MNRAS.362..505C}
\bibinfo{author}{\bibfnamefont{S.}~\bibnamefont{{Cole}}},
  \bibinfo{author}{\bibfnamefont{W.~J.} \bibnamefont{{Percival}}},
  \bibinfo{author}{\bibfnamefont{J.~A.} \bibnamefont{{Peacock}}},
  \bibinfo{author}{\bibfnamefont{P.}~\bibnamefont{{Norberg}}},
  \bibinfo{author}{\bibfnamefont{C.~M.} \bibnamefont{{Baugh}}},
  \bibinfo{author}{\bibfnamefont{C.~S.} \bibnamefont{{Frenk}}},
  \bibinfo{author}{\bibfnamefont{I.}~\bibnamefont{{Baldry}}},
  \bibinfo{author}{\bibfnamefont{J.}~\bibnamefont{{Bland-Hawthorn}}},
  \bibinfo{author}{\bibfnamefont{T.}~\bibnamefont{{Bridges}}},
  \bibinfo{author}{\bibfnamefont{R.}~\bibnamefont{{Cannon}}},
  \bibnamefont{et~al.}, \bibinfo{journal}{\mnras}
  \textbf{\bibinfo{volume}{362}}, \bibinfo{pages}{505} (\bibinfo{year}{2005}),
  \eprint{arXiv:astro-ph/0501174}.

\bibitem[{\citenamefont{{Tegmark} et~al.}(2006)\citenamefont{{Tegmark},
  {Eisenstein}, {Strauss}, {Weinberg}, {Blanton}, {Frieman}, {Fukugita},
  {Gunn}, {Hamilton}, {Knapp} et~al.}}]{2006PhRvD..74l3507T}
\bibinfo{author}{\bibfnamefont{M.}~\bibnamefont{{Tegmark}}},
  \bibinfo{author}{\bibfnamefont{D.~J.} \bibnamefont{{Eisenstein}}},
  \bibinfo{author}{\bibfnamefont{M.~A.} \bibnamefont{{Strauss}}},
  \bibinfo{author}{\bibfnamefont{D.~H.} \bibnamefont{{Weinberg}}},
  \bibinfo{author}{\bibfnamefont{M.~R.} \bibnamefont{{Blanton}}},
  \bibinfo{author}{\bibfnamefont{J.~A.} \bibnamefont{{Frieman}}},
  \bibinfo{author}{\bibfnamefont{M.}~\bibnamefont{{Fukugita}}},
  \bibinfo{author}{\bibfnamefont{J.~E.} \bibnamefont{{Gunn}}},
  \bibinfo{author}{\bibfnamefont{A.~J.~S.} \bibnamefont{{Hamilton}}},
  \bibinfo{author}{\bibfnamefont{G.~R.} \bibnamefont{{Knapp}}},
  \bibnamefont{et~al.}, \bibinfo{journal}{\prd} \textbf{\bibinfo{volume}{74}},
  \bibinfo{pages}{123507} (\bibinfo{year}{2006}),
  \eprint{arXiv:astro-ph/0608632}.

\bibitem[{\citenamefont{{Percival} et~al.}(2007)\citenamefont{{Percival},
  {Cole}, {Eisenstein}, {Nichol}, {Peacock}, {Pope}, and
  {Szalay}}}]{2007MNRAS.381.1053P}
\bibinfo{author}{\bibfnamefont{W.~J.} \bibnamefont{{Percival}}},
  \bibinfo{author}{\bibfnamefont{S.}~\bibnamefont{{Cole}}},
  \bibinfo{author}{\bibfnamefont{D.~J.} \bibnamefont{{Eisenstein}}},
  \bibinfo{author}{\bibfnamefont{R.~C.} \bibnamefont{{Nichol}}},
  \bibinfo{author}{\bibfnamefont{J.~A.} \bibnamefont{{Peacock}}},
  \bibinfo{author}{\bibfnamefont{A.~C.} \bibnamefont{{Pope}}},
  \bibnamefont{and} \bibinfo{author}{\bibfnamefont{A.~S.}
  \bibnamefont{{Szalay}}}, \bibinfo{journal}{\mnras}
  \textbf{\bibinfo{volume}{381}}, \bibinfo{pages}{1053} (\bibinfo{year}{2007}),
  \eprint{0705.3323}.

\bibitem[{\citenamefont{{Komatsu} et~al.}(2008)\citenamefont{{Komatsu},
  {Dunkley}, {Nolta}, {Bennett}, {Gold}, {Hinshaw}, {Jarosik}, {Larson},
  {Limon}, {Page} et~al.}}]{2008arXiv0803.0547K}
\bibinfo{author}{\bibfnamefont{E.}~\bibnamefont{{Komatsu}}},
  \bibinfo{author}{\bibfnamefont{J.}~\bibnamefont{{Dunkley}}},
  \bibinfo{author}{\bibfnamefont{M.~R.} \bibnamefont{{Nolta}}},
  \bibinfo{author}{\bibfnamefont{C.~L.} \bibnamefont{{Bennett}}},
  \bibinfo{author}{\bibfnamefont{B.}~\bibnamefont{{Gold}}},
  \bibinfo{author}{\bibfnamefont{G.}~\bibnamefont{{Hinshaw}}},
  \bibinfo{author}{\bibfnamefont{N.}~\bibnamefont{{Jarosik}}},
  \bibinfo{author}{\bibfnamefont{D.}~\bibnamefont{{Larson}}},
  \bibinfo{author}{\bibfnamefont{M.}~\bibnamefont{{Limon}}},
  \bibinfo{author}{\bibfnamefont{L.}~\bibnamefont{{Page}}},
  \bibnamefont{et~al.}, \bibinfo{journal}{ArXiv e-prints}
  (\bibinfo{year}{2008}), \eprint{0803.0547}.

\bibitem[{\citenamefont{Dunkley et~al.}(2008)}]{Dunkley:2008ie}
\bibinfo{author}{\bibfnamefont{J.}~\bibnamefont{Dunkley}} \bibnamefont{et~al.}
  (\bibinfo{collaboration}{WMAP}) (\bibinfo{year}{2008}), \eprint{0803.0586}.

\bibitem[{\citenamefont{{Oguri} et~al.}(2008)\citenamefont{{Oguri}, {Inada},
  {Strauss}, {Kochanek}, {Richards}, {Schneider}, {Becker}, {Fukugita},
  {Gregg}, {Hall} et~al.}}]{2008AJ....135..512O}
\bibinfo{author}{\bibfnamefont{M.}~\bibnamefont{{Oguri}}},
  \bibinfo{author}{\bibfnamefont{N.}~\bibnamefont{{Inada}}},
  \bibinfo{author}{\bibfnamefont{M.~A.} \bibnamefont{{Strauss}}},
  \bibinfo{author}{\bibfnamefont{C.~S.} \bibnamefont{{Kochanek}}},
  \bibinfo{author}{\bibfnamefont{G.~T.} \bibnamefont{{Richards}}},
  \bibinfo{author}{\bibfnamefont{D.~P.} \bibnamefont{{Schneider}}},
  \bibinfo{author}{\bibfnamefont{R.~H.} \bibnamefont{{Becker}}},
  \bibinfo{author}{\bibfnamefont{M.}~\bibnamefont{{Fukugita}}},
  \bibinfo{author}{\bibfnamefont{M.~D.} \bibnamefont{{Gregg}}},
  \bibinfo{author}{\bibfnamefont{P.~B.} \bibnamefont{{Hall}}},
  \bibnamefont{et~al.}, \bibinfo{journal}{\aj} \textbf{\bibinfo{volume}{135}},
  \bibinfo{pages}{512} (\bibinfo{year}{2008}), \eprint{0708.0825}.

\bibitem[{\citenamefont{{Kowalski} et~al.}(2008)\citenamefont{{Kowalski},
  {Rubin}, {Aldering}, {Agostinho}, {Amadon}, {Amanullah}, {Balland},
  {Barbary}, {Blanc}, {Challis} et~al.}}]{2008ApJ...686..749K}
\bibinfo{author}{\bibfnamefont{M.}~\bibnamefont{{Kowalski}}},
  \bibinfo{author}{\bibfnamefont{D.}~\bibnamefont{{Rubin}}},
  \bibinfo{author}{\bibfnamefont{G.}~\bibnamefont{{Aldering}}},
  \bibinfo{author}{\bibfnamefont{R.~J.} \bibnamefont{{Agostinho}}},
  \bibinfo{author}{\bibfnamefont{A.}~\bibnamefont{{Amadon}}},
  \bibinfo{author}{\bibfnamefont{R.}~\bibnamefont{{Amanullah}}},
  \bibinfo{author}{\bibfnamefont{C.}~\bibnamefont{{Balland}}},
  \bibinfo{author}{\bibfnamefont{K.}~\bibnamefont{{Barbary}}},
  \bibinfo{author}{\bibfnamefont{G.}~\bibnamefont{{Blanc}}},
  \bibinfo{author}{\bibfnamefont{P.~J.} \bibnamefont{{Challis}}},
  \bibnamefont{et~al.}, \bibinfo{journal}{\apj} \textbf{\bibinfo{volume}{686}},
  \bibinfo{pages}{749} (\bibinfo{year}{2008}), \eprint{0804.4142}.

\bibitem[{\citenamefont{{Albrecht} et~al.}(2006)\citenamefont{{Albrecht},
  {Bernstein}, {Cahn}, {Freedman}, {Hewitt}, {Hu}, {Huth}, {Kamionkowski},
  {Kolb}, {Knox} et~al.}}]{2006astro.ph..9591A}
\bibinfo{author}{\bibfnamefont{A.}~\bibnamefont{{Albrecht}}},
  \bibinfo{author}{\bibfnamefont{G.}~\bibnamefont{{Bernstein}}},
  \bibinfo{author}{\bibfnamefont{R.}~\bibnamefont{{Cahn}}},
  \bibinfo{author}{\bibfnamefont{W.~L.} \bibnamefont{{Freedman}}},
  \bibinfo{author}{\bibfnamefont{J.}~\bibnamefont{{Hewitt}}},
  \bibinfo{author}{\bibfnamefont{W.}~\bibnamefont{{Hu}}},
  \bibinfo{author}{\bibfnamefont{J.}~\bibnamefont{{Huth}}},
  \bibinfo{author}{\bibfnamefont{M.}~\bibnamefont{{Kamionkowski}}},
  \bibinfo{author}{\bibfnamefont{E.~W.} \bibnamefont{{Kolb}}},
  \bibinfo{author}{\bibfnamefont{L.}~\bibnamefont{{Knox}}},
  \bibnamefont{et~al.}, \bibinfo{journal}{ArXiv Astrophysics e-prints}
  (\bibinfo{year}{2006}), \eprint{arXiv:astro-ph/0609591}.

\bibitem[{\citenamefont{{Chevallier} and
  {Polarski}}(2001)}]{2001IJMPD..10..213C}
\bibinfo{author}{\bibfnamefont{M.}~\bibnamefont{{Chevallier}}}
  \bibnamefont{and}
  \bibinfo{author}{\bibfnamefont{D.}~\bibnamefont{{Polarski}}},
  \bibinfo{journal}{International Journal of Modern Physics D}
  \textbf{\bibinfo{volume}{10}}, \bibinfo{pages}{213} (\bibinfo{year}{2001}),
  \eprint{arXiv:gr-qc/0009008}.

\bibitem[{\citenamefont{{Linder}}(2003)}]{2003PhRvL..90i1301L}
\bibinfo{author}{\bibfnamefont{E.~V.} \bibnamefont{{Linder}}},
  \bibinfo{journal}{Physical Review Letters} \textbf{\bibinfo{volume}{90}},
  \bibinfo{pages}{091301} (\bibinfo{year}{2003}),
  \eprint{arXiv:astro-ph/0208512}.

\bibitem[{\citenamefont{{Tegmark}
  et~al.}(1998{\natexlab{a}})\citenamefont{{Tegmark}, {Eisenstein}, {Hu}, and
  {Kron}}}]{1998astro.ph..5117T}
\bibinfo{author}{\bibfnamefont{M.}~\bibnamefont{{Tegmark}}},
  \bibinfo{author}{\bibfnamefont{D.~J.} \bibnamefont{{Eisenstein}}},
  \bibinfo{author}{\bibfnamefont{W.}~\bibnamefont{{Hu}}}, \bibnamefont{and}
  \bibinfo{author}{\bibfnamefont{R.}~\bibnamefont{{Kron}}},
  \bibinfo{journal}{ArXiv Astrophysics e-prints}
  (\bibinfo{year}{1998}{\natexlab{a}}), \eprint{arXiv:astro-ph/9805117}.

\bibitem[{\citenamefont{{Tegmark}
  et~al.}(1998{\natexlab{b}})\citenamefont{{Tegmark}, {Eisenstein}, and
  {Hu}}}]{1998astro.ph..4168T}
\bibinfo{author}{\bibfnamefont{M.}~\bibnamefont{{Tegmark}}},
  \bibinfo{author}{\bibfnamefont{D.~J.} \bibnamefont{{Eisenstein}}},
  \bibnamefont{and} \bibinfo{author}{\bibfnamefont{W.}~\bibnamefont{{Hu}}},
  \bibinfo{journal}{ArXiv Astrophysics e-prints}
  (\bibinfo{year}{1998}{\natexlab{b}}), \eprint{arXiv:astro-ph/9804168}.

\bibitem[{\citenamefont{{Eisenstein} et~al.}(1999)\citenamefont{{Eisenstein},
  {Hu}, and {Tegmark}}}]{1999ApJ...518....2E}
\bibinfo{author}{\bibfnamefont{D.~J.} \bibnamefont{{Eisenstein}}},
  \bibinfo{author}{\bibfnamefont{W.}~\bibnamefont{{Hu}}}, \bibnamefont{and}
  \bibinfo{author}{\bibfnamefont{M.}~\bibnamefont{{Tegmark}}},
  \bibinfo{journal}{\apj} \textbf{\bibinfo{volume}{518}}, \bibinfo{pages}{2}
  (\bibinfo{year}{1999}), \eprint{arXiv:astro-ph/9807130}.

\bibitem[{\citenamefont{{Frieman} et~al.}(2003)\citenamefont{{Frieman},
  {Huterer}, {Linder}, and {Turner}}}]{2003PhRvD..67h3505F}
\bibinfo{author}{\bibfnamefont{J.~A.} \bibnamefont{{Frieman}}},
  \bibinfo{author}{\bibfnamefont{D.}~\bibnamefont{{Huterer}}},
  \bibinfo{author}{\bibfnamefont{E.~V.} \bibnamefont{{Linder}}},
  \bibnamefont{and} \bibinfo{author}{\bibfnamefont{M.~S.}
  \bibnamefont{{Turner}}}, \bibinfo{journal}{\prd}
  \textbf{\bibinfo{volume}{67}}, \bibinfo{pages}{083505}
  (\bibinfo{year}{2003}), \eprint{arXiv:astro-ph/0208100}.

\bibitem[{\citenamefont{{Cole} et~al.}(2006)\citenamefont{{Cole}, {Sanchez},
  and {Wilkins}}}]{2006astro.ph.11178C}
\bibinfo{author}{\bibfnamefont{S.}~\bibnamefont{{Cole}}},
  \bibinfo{author}{\bibfnamefont{A.~G.} \bibnamefont{{Sanchez}}},
  \bibnamefont{and}
  \bibinfo{author}{\bibfnamefont{S.}~\bibnamefont{{Wilkins}}},
  \bibinfo{journal}{ArXiv Astrophysics e-prints}  (\bibinfo{year}{2006}),
  \eprint{arXiv:astro-ph/0611178}.

\bibitem[{\citenamefont{{Wang}}(2008)}]{2008PhRvD..77l3525W}
\bibinfo{author}{\bibfnamefont{Y.}~\bibnamefont{{Wang}}},
  \bibinfo{journal}{\prd} \textbf{\bibinfo{volume}{77}},
  \bibinfo{pages}{123525} (\bibinfo{year}{2008}), \eprint{0803.4295}.

\bibitem[{\citenamefont{{Xia} et~al.}(2008)\citenamefont{{Xia}, {Li}, {Zhao},
  and {Zhang}}}]{2008PhRvD..78h3524X}
\bibinfo{author}{\bibfnamefont{J.-Q.} \bibnamefont{{Xia}}},
  \bibinfo{author}{\bibfnamefont{H.}~\bibnamefont{{Li}}},
  \bibinfo{author}{\bibfnamefont{G.-B.} \bibnamefont{{Zhao}}},
  \bibnamefont{and} \bibinfo{author}{\bibfnamefont{X.}~\bibnamefont{{Zhang}}},
  \bibinfo{journal}{\prd} \textbf{\bibinfo{volume}{78}},
  \bibinfo{pages}{083524} (\bibinfo{year}{2008}), \eprint{0807.3878}.

\bibitem[{\citenamefont{{Biswas} and {Wandelt}}(2009)}]{2009arXiv0903.2532B}
\bibinfo{author}{\bibfnamefont{R.}~\bibnamefont{{Biswas}}} \bibnamefont{and}
  \bibinfo{author}{\bibfnamefont{B.~D.} \bibnamefont{{Wandelt}}},
  \bibinfo{journal}{ArXiv e-prints}  (\bibinfo{year}{2009}),
  \eprint{0903.2532}.

\bibitem[{\citenamefont{{Ho} et~al.}(2006)\citenamefont{{Ho}, {Bahcall}, and
  {Bode}}}]{2006ApJ...647....8H}
\bibinfo{author}{\bibfnamefont{S.}~\bibnamefont{{Ho}}},
  \bibinfo{author}{\bibfnamefont{N.}~\bibnamefont{{Bahcall}}},
  \bibnamefont{and} \bibinfo{author}{\bibfnamefont{P.}~\bibnamefont{{Bode}}},
  \bibinfo{journal}{\apj} \textbf{\bibinfo{volume}{647}}, \bibinfo{pages}{8}
  (\bibinfo{year}{2006}), \eprint{arXiv:astro-ph/0511776}.

\bibitem[{\citenamefont{{Park} and {Lee}}(2007)}]{2007PhRvL..98h1301P}
\bibinfo{author}{\bibfnamefont{D.}~\bibnamefont{{Park}}} \bibnamefont{and}
  \bibinfo{author}{\bibfnamefont{J.}~\bibnamefont{{Lee}}},
  \bibinfo{journal}{Physical Review Letters} \textbf{\bibinfo{volume}{98}},
  \bibinfo{pages}{081301} (\bibinfo{year}{2007}).

\bibitem[{\citenamefont{{Lee} and {Park}}(2007)}]{2007arXiv0704.0881L}
\bibinfo{author}{\bibfnamefont{J.}~\bibnamefont{{Lee}}} \bibnamefont{and}
  \bibinfo{author}{\bibfnamefont{D.}~\bibnamefont{{Park}}},
  \bibinfo{journal}{ArXiv e-prints}  (\bibinfo{year}{2007}),
  \eprint{0704.0881}.

\bibitem[{\citenamefont{{El-Ad} and {Piran}}(1997)}]{1997ApJ...491..421E}
\bibinfo{author}{\bibfnamefont{H.}~\bibnamefont{{El-Ad}}} \bibnamefont{and}
  \bibinfo{author}{\bibfnamefont{T.}~\bibnamefont{{Piran}}},
  \bibinfo{journal}{\apj} \textbf{\bibinfo{volume}{491}}, \bibinfo{pages}{421}
  (\bibinfo{year}{1997}), \eprint{arXiv:astro-ph/9702135}.

\bibitem[{\citenamefont{{Hoyle} and {Vogeley}}(2001)}]{2001astro.ph.10449H}
\bibinfo{author}{\bibfnamefont{F.}~\bibnamefont{{Hoyle}}} \bibnamefont{and}
  \bibinfo{author}{\bibfnamefont{M.~S.} \bibnamefont{{Vogeley}}},
  \bibinfo{journal}{ArXiv Astrophysics e-prints}  (\bibinfo{year}{2001}),
  \eprint{arXiv:astro-ph/0110449}.

\bibitem[{\citenamefont{{Hoyle} and {Vogeley}}(2002)}]{2002ApJ...566..641H}
\bibinfo{author}{\bibfnamefont{F.}~\bibnamefont{{Hoyle}}} \bibnamefont{and}
  \bibinfo{author}{\bibfnamefont{M.~S.} \bibnamefont{{Vogeley}}},
  \bibinfo{journal}{\apj} \textbf{\bibinfo{volume}{566}}, \bibinfo{pages}{641}
  (\bibinfo{year}{2002}), \eprint{arXiv:astro-ph/0109357}.

\bibitem[{\citenamefont{{Neyrinck}}(2008)}]{2008MNRAS.386.2101N}
\bibinfo{author}{\bibfnamefont{M.~C.} \bibnamefont{{Neyrinck}}},
  \bibinfo{journal}{\mnras} \textbf{\bibinfo{volume}{386}},
  \bibinfo{pages}{2101} (\bibinfo{year}{2008}), \eprint{0712.3049}.

\bibitem[{\citenamefont{{Colberg} et~al.}(2008)\citenamefont{{Colberg},
  {Pearce}, {Foster}, {Platen}, {Brunino}, {Neyrinck}, {Basilakos}, {Fairall},
  {Feldman}, {Gottl{\"o}ber} et~al.}}]{2008MNRAS.387..933C}
\bibinfo{author}{\bibfnamefont{J.~M.} \bibnamefont{{Colberg}}},
  \bibinfo{author}{\bibfnamefont{F.}~\bibnamefont{{Pearce}}},
  \bibinfo{author}{\bibfnamefont{C.}~\bibnamefont{{Foster}}},
  \bibinfo{author}{\bibfnamefont{E.}~\bibnamefont{{Platen}}},
  \bibinfo{author}{\bibfnamefont{R.}~\bibnamefont{{Brunino}}},
  \bibinfo{author}{\bibfnamefont{M.}~\bibnamefont{{Neyrinck}}},
  \bibinfo{author}{\bibfnamefont{S.}~\bibnamefont{{Basilakos}}},
  \bibinfo{author}{\bibfnamefont{A.}~\bibnamefont{{Fairall}}},
  \bibinfo{author}{\bibfnamefont{H.}~\bibnamefont{{Feldman}}},
  \bibinfo{author}{\bibfnamefont{S.}~\bibnamefont{{Gottl{\"o}ber}}},
  \bibnamefont{et~al.}, \bibinfo{journal}{\mnras}
  \textbf{\bibinfo{volume}{387}}, \bibinfo{pages}{933} (\bibinfo{year}{2008}),
  \eprint{0803.0918}.

\bibitem[{\citenamefont{{Hoyle} and {Vogeley}}(2004)}]{2004ApJ...607..751H}
\bibinfo{author}{\bibfnamefont{F.}~\bibnamefont{{Hoyle}}} \bibnamefont{and}
  \bibinfo{author}{\bibfnamefont{M.~S.} \bibnamefont{{Vogeley}}},
  \bibinfo{journal}{\apj} \textbf{\bibinfo{volume}{607}}, \bibinfo{pages}{751}
  (\bibinfo{year}{2004}), \eprint{arXiv:astro-ph/0312533}.

\bibitem[{\citenamefont{{Goldberg} et~al.}(2005)\citenamefont{{Goldberg},
  {Jones}, {Hoyle}, {Rojas}, {Vogeley}, and {Blanton}}}]{2005ApJ...621..643G}
\bibinfo{author}{\bibfnamefont{D.~M.} \bibnamefont{{Goldberg}}},
  \bibinfo{author}{\bibfnamefont{T.~D.} \bibnamefont{{Jones}}},
  \bibinfo{author}{\bibfnamefont{F.}~\bibnamefont{{Hoyle}}},
  \bibinfo{author}{\bibfnamefont{R.~R.} \bibnamefont{{Rojas}}},
  \bibinfo{author}{\bibfnamefont{M.~S.} \bibnamefont{{Vogeley}}},
  \bibnamefont{and} \bibinfo{author}{\bibfnamefont{M.~R.}
  \bibnamefont{{Blanton}}}, \bibinfo{journal}{\apj}
  \textbf{\bibinfo{volume}{621}}, \bibinfo{pages}{643} (\bibinfo{year}{2005}),
  \eprint{arXiv:astro-ph/0406527}.

\bibitem[{\citenamefont{{Tikhonov}}(2007)}]{2007AstL...33..499T}
\bibinfo{author}{\bibfnamefont{A.~V.} \bibnamefont{{Tikhonov}}},
  \bibinfo{journal}{Astronomy Letters} \textbf{\bibinfo{volume}{33}},
  \bibinfo{pages}{499} (\bibinfo{year}{2007}), \eprint{0707.4283}.

\bibitem[{\citenamefont{{Foster} and {Nelson}}(2009)}]{2009arXiv0904.4721F}
\bibinfo{author}{\bibfnamefont{C.}~\bibnamefont{{Foster}}} \bibnamefont{and}
  \bibinfo{author}{\bibfnamefont{L.~A.} \bibnamefont{{Nelson}}},
  \bibinfo{journal}{ArXiv e-prints}  (\bibinfo{year}{2009}),
  \eprint{0904.4721}.

\bibitem[{\citenamefont{{Sheth} et~al.}(2001)\citenamefont{{Sheth}, {Mo}, and
  {Tormen}}}]{2001MNRAS.323....1S}
\bibinfo{author}{\bibfnamefont{R.~K.} \bibnamefont{{Sheth}}},
  \bibinfo{author}{\bibfnamefont{H.~J.} \bibnamefont{{Mo}}}, \bibnamefont{and}
  \bibinfo{author}{\bibfnamefont{G.}~\bibnamefont{{Tormen}}},
  \bibinfo{journal}{\mnras} \textbf{\bibinfo{volume}{323}}, \bibinfo{pages}{1}
  (\bibinfo{year}{2001}), \eprint{arXiv:astro-ph/9907024}.

\bibitem[{\citenamefont{{Sheth} and {Tormen}}(2002)}]{2002MNRAS.329...61S}
\bibinfo{author}{\bibfnamefont{R.~K.} \bibnamefont{{Sheth}}} \bibnamefont{and}
  \bibinfo{author}{\bibfnamefont{G.}~\bibnamefont{{Tormen}}},
  \bibinfo{journal}{\mnras} \textbf{\bibinfo{volume}{329}}, \bibinfo{pages}{61}
  (\bibinfo{year}{2002}), \eprint{arXiv:astro-ph/0105113}.

\bibitem[{\citenamefont{{Chiueh} and {Lee}}(2001)}]{2001ApJ...555...83C}
\bibinfo{author}{\bibfnamefont{T.}~\bibnamefont{{Chiueh}}} \bibnamefont{and}
  \bibinfo{author}{\bibfnamefont{J.}~\bibnamefont{{Lee}}},
  \bibinfo{journal}{\apj} \textbf{\bibinfo{volume}{555}}, \bibinfo{pages}{83}
  (\bibinfo{year}{2001}), \eprint{arXiv:astro-ph/0010286}.

\bibitem[{\citenamefont{{van de Weygaert} et~al.}(2004)\citenamefont{{van de
  Weygaert}, {Sheth}, and {Platen}}}]{2004ogci.conf...58V}
\bibinfo{author}{\bibfnamefont{R.}~\bibnamefont{{van de Weygaert}}},
  \bibinfo{author}{\bibfnamefont{R.}~\bibnamefont{{Sheth}}}, \bibnamefont{and}
  \bibinfo{author}{\bibfnamefont{E.}~\bibnamefont{{Platen}}}, in
  \emph{\bibinfo{booktitle}{IAU Colloq. 195: Outskirts of Galaxy Clusters:
  Intense Life in the Suburbs}}, edited by
  \bibinfo{editor}{\bibfnamefont{A.}~\bibnamefont{{Diaferio}}}
  (\bibinfo{year}{2004}), pp. \bibinfo{pages}{58--63}.

\bibitem[{\citenamefont{{Shandarin} et~al.}(2006)\citenamefont{{Shandarin},
  {Feldman}, {Heitmann}, and {Habib}}}]{2006MNRAS.367.1629S}
\bibinfo{author}{\bibfnamefont{S.}~\bibnamefont{{Shandarin}}},
  \bibinfo{author}{\bibfnamefont{H.~A.} \bibnamefont{{Feldman}}},
  \bibinfo{author}{\bibfnamefont{K.}~\bibnamefont{{Heitmann}}},
  \bibnamefont{and} \bibinfo{author}{\bibfnamefont{S.}~\bibnamefont{{Habib}}},
  \bibinfo{journal}{\mnras} \textbf{\bibinfo{volume}{367}},
  \bibinfo{pages}{1629} (\bibinfo{year}{2006}),
  \eprint{arXiv:astro-ph/0509858}.

\bibitem[{\citenamefont{{Lavaux} and {Wandelt}}(2009)}]{2009arXiv0906.4101L}
\bibinfo{author}{\bibfnamefont{G.}~\bibnamefont{{Lavaux}}} \bibnamefont{and}
  \bibinfo{author}{\bibfnamefont{B.~D.} \bibnamefont{{Wandelt}}},
  \bibinfo{journal}{ArXiv e-prints}  (\bibinfo{year}{2009}),
  \eprint{0906.4101}.

\bibitem[{\citenamefont{{Bardeen} et~al.}(1986)\citenamefont{{Bardeen}, {Bond},
  {Kaiser}, and {Szalay}}}]{1986ApJ...304...15B}
\bibinfo{author}{\bibfnamefont{J.~M.} \bibnamefont{{Bardeen}}},
  \bibinfo{author}{\bibfnamefont{J.~R.} \bibnamefont{{Bond}}},
  \bibinfo{author}{\bibfnamefont{N.}~\bibnamefont{{Kaiser}}}, \bibnamefont{and}
  \bibinfo{author}{\bibfnamefont{A.~S.} \bibnamefont{{Szalay}}},
  \bibinfo{journal}{\apj} \textbf{\bibinfo{volume}{304}}, \bibinfo{pages}{15}
  (\bibinfo{year}{1986}).

\bibitem[{\citenamefont{{Hahn} et~al.}(2007)\citenamefont{{Hahn}, {Porciani},
  {Carollo}, and {Dekel}}}]{2007MNRAS.375..489H}
\bibinfo{author}{\bibfnamefont{O.}~\bibnamefont{{Hahn}}},
  \bibinfo{author}{\bibfnamefont{C.}~\bibnamefont{{Porciani}}},
  \bibinfo{author}{\bibfnamefont{C.~M.} \bibnamefont{{Carollo}}},
  \bibnamefont{and} \bibinfo{author}{\bibfnamefont{A.}~\bibnamefont{{Dekel}}},
  \bibinfo{journal}{\mnras} \textbf{\bibinfo{volume}{375}},
  \bibinfo{pages}{489} (\bibinfo{year}{2007}), \eprint{arXiv:astro-ph/0610280}.

\bibitem[{\citenamefont{{Bond} and {Myers}}(1996)}]{1996ApJS..103....1B}
\bibinfo{author}{\bibfnamefont{J.~R.} \bibnamefont{{Bond}}} \bibnamefont{and}
  \bibinfo{author}{\bibfnamefont{S.~T.} \bibnamefont{{Myers}}},
  \bibinfo{journal}{\apjs} \textbf{\bibinfo{volume}{103}}, \bibinfo{pages}{1}
  (\bibinfo{year}{1996}).

\bibitem[{\citenamefont{{Percival}}(2005)}]{2005A&A...443..819P}
\bibinfo{author}{\bibfnamefont{W.~J.} \bibnamefont{{Percival}}},
  \bibinfo{journal}{\aap} \textbf{\bibinfo{volume}{443}}, \bibinfo{pages}{819}
  (\bibinfo{year}{2005}), \eprint{arXiv:astro-ph/0508156}.

\bibitem[{\citenamefont{{Basilakos}}(2003)}]{2003ApJ...590..636B}
\bibinfo{author}{\bibfnamefont{S.}~\bibnamefont{{Basilakos}}},
  \bibinfo{journal}{\apj} \textbf{\bibinfo{volume}{590}}, \bibinfo{pages}{636}
  (\bibinfo{year}{2003}), \eprint{arXiv:astro-ph/0303112}.

\bibitem[{\citenamefont{{White} and {Silk}}(1979)}]{1979ApJ...231....1W}
\bibinfo{author}{\bibfnamefont{S.~D.~M.} \bibnamefont{{White}}}
  \bibnamefont{and} \bibinfo{author}{\bibfnamefont{J.}~\bibnamefont{{Silk}}},
  \bibinfo{journal}{\apj} \textbf{\bibinfo{volume}{231}}, \bibinfo{pages}{1}
  (\bibinfo{year}{1979}).

\bibitem[{\citenamefont{{Icke}}(1984)}]{1984MNRAS.206P...1I}
\bibinfo{author}{\bibfnamefont{V.}~\bibnamefont{{Icke}}},
  \bibinfo{journal}{\mnras} \textbf{\bibinfo{volume}{206}}, \bibinfo{pages}{1P}
  (\bibinfo{year}{1984}).

\bibitem[{\citenamefont{{van de Weygaert} and
  {Bertschinger}}(1996)}]{1996MNRAS.281...84V}
\bibinfo{author}{\bibfnamefont{R.}~\bibnamefont{{van de Weygaert}}}
  \bibnamefont{and}
  \bibinfo{author}{\bibfnamefont{E.}~\bibnamefont{{Bertschinger}}},
  \bibinfo{journal}{\mnras} \textbf{\bibinfo{volume}{281}}, \bibinfo{pages}{84}
  (\bibinfo{year}{1996}), \eprint{arXiv:astro-ph/9507024}.

\bibitem[{\citenamefont{{Lewis} et~al.}(2000)\citenamefont{{Lewis},
  {Challinor}, and {Lasenby}}}]{2000ApJ...538..473L}
\bibinfo{author}{\bibfnamefont{A.}~\bibnamefont{{Lewis}}},
  \bibinfo{author}{\bibfnamefont{A.}~\bibnamefont{{Challinor}}},
  \bibnamefont{and}
  \bibinfo{author}{\bibfnamefont{A.}~\bibnamefont{{Lasenby}}},
  \bibinfo{journal}{\apj} \textbf{\bibinfo{volume}{538}}, \bibinfo{pages}{473}
  (\bibinfo{year}{2000}), \eprint{arXiv:astro-ph/9911177}.

\bibitem[{\citenamefont{{Mo} and {White}}(1996)}]{1996MNRAS.282..347M}
\bibinfo{author}{\bibfnamefont{H.~J.} \bibnamefont{{Mo}}} \bibnamefont{and}
  \bibinfo{author}{\bibfnamefont{S.~D.~M.} \bibnamefont{{White}}},
  \bibinfo{journal}{\mnras} \textbf{\bibinfo{volume}{282}},
  \bibinfo{pages}{347} (\bibinfo{year}{1996}), \eprint{arXiv:astro-ph/9512127}.

\bibitem[{\citenamefont{{Sheth} and {van de
  Weygaert}}(2004)}]{2004MNRAS.350..517S}
\bibinfo{author}{\bibfnamefont{R.~K.} \bibnamefont{{Sheth}}} \bibnamefont{and}
  \bibinfo{author}{\bibfnamefont{R.}~\bibnamefont{{van de Weygaert}}},
  \bibinfo{journal}{\mnras} \textbf{\bibinfo{volume}{350}},
  \bibinfo{pages}{517} (\bibinfo{year}{2004}), \eprint{arXiv:astro-ph/0311260}.

\bibitem[{\citenamefont{{Press} and {Schechter}}(1974)}]{1974ApJ...187..425P}
\bibinfo{author}{\bibfnamefont{W.~H.} \bibnamefont{{Press}}} \bibnamefont{and}
  \bibinfo{author}{\bibfnamefont{P.}~\bibnamefont{{Schechter}}},
  \bibinfo{journal}{\apj} \textbf{\bibinfo{volume}{187}}, \bibinfo{pages}{425}
  (\bibinfo{year}{1974}).

\bibitem[{\citenamefont{{Blanton} et~al.}(2001)\citenamefont{{Blanton},
  {Dalcanton}, {Eisenstein}, {Loveday}, {Strauss}, {SubbaRao}, {Weinberg},
  {Anderson}, {Annis}, {Bahcall} et~al.}}]{2001AJ....121.2358B}
\bibinfo{author}{\bibfnamefont{M.~R.} \bibnamefont{{Blanton}}},
  \bibinfo{author}{\bibfnamefont{J.}~\bibnamefont{{Dalcanton}}},
  \bibinfo{author}{\bibfnamefont{D.}~\bibnamefont{{Eisenstein}}},
  \bibinfo{author}{\bibfnamefont{J.}~\bibnamefont{{Loveday}}},
  \bibinfo{author}{\bibfnamefont{M.~A.} \bibnamefont{{Strauss}}},
  \bibinfo{author}{\bibfnamefont{M.}~\bibnamefont{{SubbaRao}}},
  \bibinfo{author}{\bibfnamefont{D.~H.} \bibnamefont{{Weinberg}}},
  \bibinfo{author}{\bibfnamefont{J.~E.} \bibnamefont{{Anderson}},
  \bibfnamefont{Jr.}},
  \bibinfo{author}{\bibfnamefont{J.}~\bibnamefont{{Annis}}},
  \bibinfo{author}{\bibfnamefont{N.~A.} \bibnamefont{{Bahcall}}},
  \bibnamefont{et~al.}, \bibinfo{journal}{\aj} \textbf{\bibinfo{volume}{121}},
  \bibinfo{pages}{2358} (\bibinfo{year}{2001}).

\bibitem[{\citenamefont{{The Planck
  Collaboration}}(2006)}]{2006astro.ph..4069T}
\bibinfo{author}{\bibnamefont{{The Planck Collaboration}}},
  \bibinfo{journal}{ArXiv Astrophysics e-prints}  (\bibinfo{year}{2006}),
  \eprint{arXiv:astro-ph/0604069}.

\bibitem[{\citenamefont{{Zhan} et~al.}(2008)\citenamefont{{Zhan}, {Wang},
  {Pinto}, and {Tyson}}}]{2008ApJ...675L...1Z}
\bibinfo{author}{\bibfnamefont{H.}~\bibnamefont{{Zhan}}},
  \bibinfo{author}{\bibfnamefont{L.}~\bibnamefont{{Wang}}},
  \bibinfo{author}{\bibfnamefont{P.}~\bibnamefont{{Pinto}}}, \bibnamefont{and}
  \bibinfo{author}{\bibfnamefont{J.~A.} \bibnamefont{{Tyson}}},
  \bibinfo{journal}{\apjl} \textbf{\bibinfo{volume}{675}}, \bibinfo{pages}{L1}
  (\bibinfo{year}{2008}), \eprint{0801.3659}.

\bibitem[{\citenamefont{{LSST Science Collaborations} and {LSST
  Project}.}(2009)}]{2009arXiv0912.0201L}
\bibinfo{author}{\bibnamefont{{LSST Science Collaborations}}} \bibnamefont{and}
  \bibinfo{author}{\bibnamefont{{LSST Project}.}}, \bibinfo{journal}{ArXiv
  e-prints}  (\bibinfo{year}{2009}), \eprint{0912.0201}.

\bibitem[{\citenamefont{{Wang} et~al.}(2004)\citenamefont{{Wang}, {Khoury},
  {Haiman}, and {May}}}]{2004PhRvD..70l3008W}
\bibinfo{author}{\bibfnamefont{S.}~\bibnamefont{{Wang}}},
  \bibinfo{author}{\bibfnamefont{J.}~\bibnamefont{{Khoury}}},
  \bibinfo{author}{\bibfnamefont{Z.}~\bibnamefont{{Haiman}}}, \bibnamefont{and}
  \bibinfo{author}{\bibfnamefont{M.}~\bibnamefont{{May}}},
  \bibinfo{journal}{\prd} \textbf{\bibinfo{volume}{70}},
  \bibinfo{pages}{123008} (\bibinfo{year}{2004}),
  \eprint{arXiv:astro-ph/0406331}.

\bibitem[{\citenamefont{{Ryden}}(1995)}]{1995ApJ...452...25R}
\bibinfo{author}{\bibfnamefont{B.~S.} \bibnamefont{{Ryden}}},
  \bibinfo{journal}{\apj} \textbf{\bibinfo{volume}{452}}, \bibinfo{pages}{25}
  (\bibinfo{year}{1995}), \eprint{arXiv:astro-ph/9506028}.

\bibitem[{\citenamefont{{Ryden} and {Melott}}(1996)}]{1996ApJ...470..160R}
\bibinfo{author}{\bibfnamefont{B.~S.} \bibnamefont{{Ryden}}} \bibnamefont{and}
  \bibinfo{author}{\bibfnamefont{A.~L.} \bibnamefont{{Melott}}},
  \bibinfo{journal}{\apj} \textbf{\bibinfo{volume}{470}}, \bibinfo{pages}{160}
  (\bibinfo{year}{1996}), \eprint{arXiv:astro-ph/9510108}.

\bibitem[{\citenamefont{{Blumenthal} et~al.}(1992)\citenamefont{{Blumenthal},
  {da Costa}, {Goldwirth}, {Lecar}, and {Piran}}}]{1992ApJ...388..234B}
\bibinfo{author}{\bibfnamefont{G.~R.} \bibnamefont{{Blumenthal}}},
  \bibinfo{author}{\bibfnamefont{L.~N.} \bibnamefont{{da Costa}}},
  \bibinfo{author}{\bibfnamefont{D.~S.} \bibnamefont{{Goldwirth}}},
  \bibinfo{author}{\bibfnamefont{M.}~\bibnamefont{{Lecar}}}, \bibnamefont{and}
  \bibinfo{author}{\bibfnamefont{T.}~\bibnamefont{{Piran}}},
  \bibinfo{journal}{\apj} \textbf{\bibinfo{volume}{388}}, \bibinfo{pages}{234}
  (\bibinfo{year}{1992}).

\bibitem[{\citenamefont{{Eisenstein} and {Loeb}}(1995)}]{1995ApJ...439..520E}
\bibinfo{author}{\bibfnamefont{D.~J.} \bibnamefont{{Eisenstein}}}
  \bibnamefont{and} \bibinfo{author}{\bibfnamefont{A.}~\bibnamefont{{Loeb}}},
  \bibinfo{journal}{\apj} \textbf{\bibinfo{volume}{439}}, \bibinfo{pages}{520}
  (\bibinfo{year}{1995}), \eprint{arXiv:astro-ph/9405012}.

\end{thebibliography}
\appendix
\section{Other Parametrizations of Asphericity of fluctuations}
\label{OtherParametrizations}
A popular choice~\citep{1986ApJ...304...15B} for density profiles, or 
~\citep{2001MNRAS.323....1S, 2002MNRAS.329...61S} expresses this in terms of
"ellipticity" and "prolateness" for the tidal ellipsoid:
\begin{equation}
\label{prolateness}
e = \frac{(\lambda_1 - \lambda_3)}{ 2 (\lambda_1 +\lambda_2+\lambda_3)}\qquad
p = \frac{(\lambda_1 + \lambda_3 -2\lambda_2)}{ 2 (\lambda_1 +\lambda_2+\lambda_3)}
\end{equation}
Since this is a function of the eigenvalues $\{\lambda\},$ there is 
a one to one correspondence with the asphericity parameters describing
the void ellipsoid in Eqn.~\ref{eqn:VoidEllipsoidEllipticity}.
\section{Generalized Excursion Set Formalism}
\label{appendix:gesf}
It should also be noted that the Doroshkevich formula is based on
conditioning on the variance within a smoothing scale $R$ at initial times (or equivalently Lagrangian smoothing
scale R), rather than the size of the structures themselves at 
later times. 
This seems suitable for void finders such as DIVA ~\citep{2009arXiv0906.4101L}, which use the variance
$\sigma_R$ as a parameter, but may be unsuitable for use with other 
void finders which find voids of particular radii at redshifts. 

In order to confront data obtained from the class of void finding 
algorithm based on clustering of underdense regions which uses the sizes 
of voids as parameters, one needs to theoretically study a distribution
of shapes of voids for different sizes. This in turn requires a 
theoretical definition of the void boundary. 
A void expands faster than the background universe, and this results in 
shell crossing forming the denser void wall. Accordingly,
~\citet{1992ApJ...388..234B}
argued that the formation of a void corresponds to this shell crossing
and is thus directly analogous to the 
collapse of a halo into a point in the spherical collapse model . 
In a spherical expansion model, they found that the linearly 
extrapolated density field at the time of shell crossing, 
is $\delta_v=-2.81$. Following them, we assume a void forms when the 
linearly extrapolated underdensity inside it reaches this critical value 
which is analogous to the critical overdensity of the Press-Schechter method. 
Since, we are interested in the asphericity of the voids, we use a 
generalized excursion set method to construct a distribution of 
the ellipticity of the tidal tensor of points at early times, given that 
they evolve to form the voids of Lagrangian size $R$. 
Each point mass belongs to a void of a certain radius at some redshift,
 and had an initial Tidal tensor. 
The goal of this method is to assign to each point mass in Lagrangian 
space (a) the radius of the void to which this point will belong at any
 given redshift, 
and (b) the eigenvalues $\{\lambda_1,\lambda_2,\lambda_3\}$ of the initial
 tidal tensor  at that point.
We provide a brief summary of the method and the results here.\\

We start with a large 
Lagrangian radius $R$, so the smoothed variance of density fluctuations 
$\sigma(R)$ is $0$ and the tidal tensor $T$ is taken to be zero. 
We perform a random walk where we decrease the radius $R$ by 
steps of $\Delta R$ (thereby increasing $\sigma(R)$ at each step), and 
execute six dimensional random walk in the independent elements of the
Tidal Tensor, with the probability of each element $P(\Delta T_{i,j},R)$ 
at radius $R$ depending on $\sigma(R)$ and equal to the probability implied
by the statistics of Gaussian fields.  The random walk 
is stopped at the largest value of $R$  when the linearized density field 
given by the sum of the eigenvalues of the tidal tensor 
$T_{i,j}(R)=\sum_{>R}\Delta T_{i,j}$
crosses the critical value of $-2.81$, and the values of the eigenvalues
$\{\lambda_1,\lambda_2,\lambda_3\}$ and $R$ at the point of termination 
are taken to be a sample of the tidal tensor eigenvalues and void size
 at that initial redshift. The mass function of voids thus obtained
is identical to the mass function that would be obtained by a spherical
collapse model (without accounting for the Void in Cloud problem), 
but it gives us a distribution of the asphericity parameters.
Repeating the process, one can construct samples
of the multivariate distribution $n(R,\{\lambda\})$, from which one can 
obtain samples for any particular asphericity parameter for voids of size 
$R$ by restricting the samples to a bin around $R$ and marginalizing over 
other the asphericity parameter.\\
In Fig.~\ref{fig:gesfcomp}, we show the histograms of the ``ellipticity" 
$e$ and ``prolateness" $p$ parameters of Eqn.~\ref{prolateness} of 
proto-voids that form voids of radius $R$ obtained from the generalized 
excursion set formalism for different values of $\nu={\delta\over \sigma}$ 
overplotted with
the distribution implied by the Doroshkevich formula in terms of the
same parameters when smoothed over a radius $R$.
\begin{figure*}
\begin{center}
\includegraphics[width=0.9\textwidth]{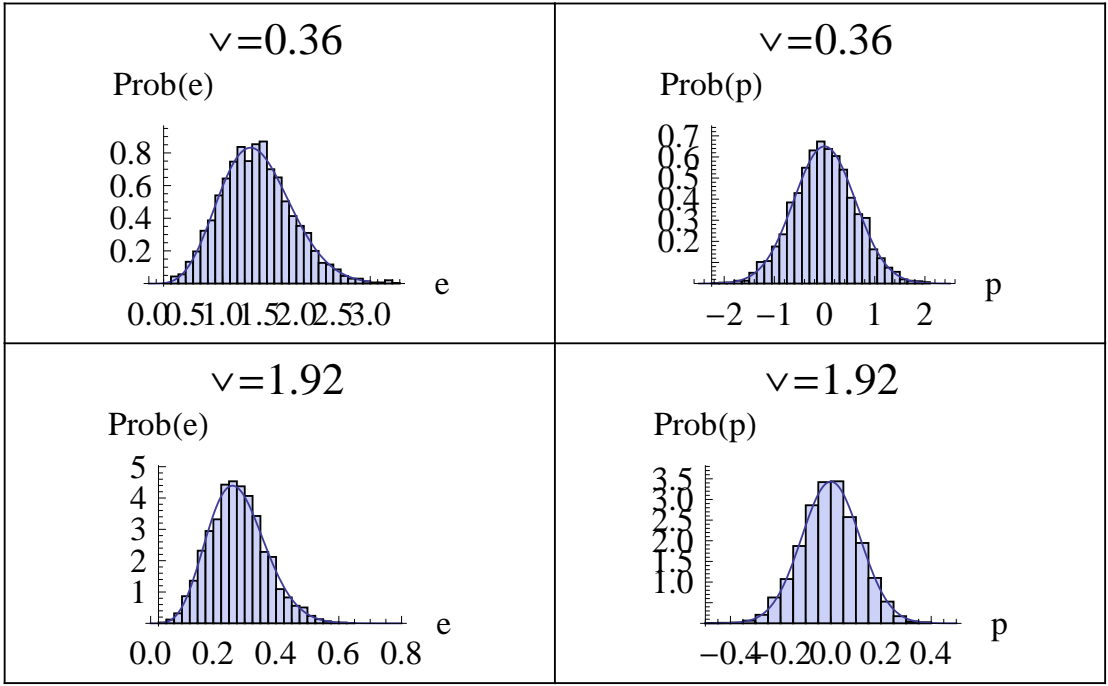}
\end{center}
\caption{Comparison of the distribution of ellipticity $e$ 
(prolateness $p$) in the left (right) panel according to Doroshkevich 
formula with the distribution obtained from generalized excursion set 
formalism described for different values of $\nu=\frac{\delta}{\sigma}$} 
\label{fig:gesfcomp}
\end{figure*}
The figure shows good agreement between the two. 
As discussed, 
in ~\citet{2009arXiv0906.4101L}, it was shown that the samples of the eigenvalues of the
tidal tensor of fields smoothed at a particular smoothing scale 
$R_{\text{Smooth}}$ are similar to the Doroshkevich Formula. Together, this
implies that both methods while giving slightly different results are 
fairly consistent, increasing our confidence in these methods.
\section{Eulerian Evolution of Triaxial Systems~}
\label{appendix:EllipsoidalCollapse}
It is well known that the Zeldovich approximation fails to describe 
structure formation at small scales in the vicinity of density peaks. 
This is because in high density regions, caustics form making the 
Lagrangian mapping non-invertible. The onset of this problem is 
characterized by shell crossing. Thus, at the minima of the density field 
well inside a void, which is our region of interest, the Zeldovich 
approximation should work well. Here, we compare with another 
approximation which works reasonably well in the vicinity of peaks. 

The evolution of a homogeneous ellipsoid of (real) density contrast 
$\delta_m(t) $
 in a homogeneous and isotropic flat LCDM universe with 
scale factor $\bar{a}$ and densities $\rho_m(t)$ and $\rho_{vac}$
has been studied
~\citep{1995ApJ...439..520E,1996ApJS..103....1B}.
The equation of motion of the scale factor, $a_i$ and $i=1,2,3$ , of three principal axes of the ellipsoid can be studied in terms of 
a second order Taylor expansion of the gravitational potential
~\citep{1996ApJS..103....1B}:
\begin{equation}
\frac{d^2}{dt^2}a_i= 
\frac{4\pi G}{3}(2 \rho_{vac} - \rho_m(t))a_i(t)
-\frac{4\pi G}{3} \rho_m(t)\delta_m(t)a_i(t) -
\frac{4\pi G}{3} \rho_m(t)(\frac{3b'_i(t)\delta_m(t)}{2}+ 3\lambda'_{ext}(t))a_i(t) 
\end{equation}
where the term in the first parenthesis is the effect of the 
usual background expansion in a flat LCDM model, the second term is the
effect of the perturbation as in spherical collapse, and the third term 
models the effect of the aspherical nature of the perturbation itself, 
and the external tides. 
The quantities $b'_i(t)$ are defined by
 \begin{equation}
 b'_j(t)=a_1(t) a_2(t) a_3(t) \int_0^{\infty} \frac{d \tau}{\left(a_j^2(t) +\tau\right) \left(a_1^2(t) +\tau\right)^{1/2} \left(a_2^2(t) +\tau\right)^{1/2} \left(a_3^2(t) +\tau\right)^{1/2} }-\frac{2}{3}
 \end{equation}
 while we use the 
the two approximation presented in \citet{1996ApJS..103....1B} for the external tidal field $\lambda^{\prime}(t)$: 
  \begin{eqnarray}
  \text{linear external tide approximation:}&& \lambda'_i(t)= \lambda_i(t)-\delta(t)/3 \label{equ:linear} \\
  \text{nonlinear external tide approximation:} && \lambda'_i(t)=5 b'_i(t)/4 \label{equ:nonlinear}
  \end{eqnarray}
  $\lambda_i$ are, as before, the eigenvalues of the tidal tensor and $\delta$ is the linearly extrapolated initial overdensity and they are proportional to the linear growth factor $D(t)$. 
 
 The initial conditions are set by using the Zeldovich approximation and are:
 
 \begin{eqnarray}
 a_i(t_{init})&=&\bar{a}(t_{init})(1-\lambda_i(t_{init})) \\
 \dot{a}_i(t_{init})&=& H(t_{init}) a_i(t_{init})-\bar{a}(t_{init}) H_D(t_{init}) \lambda_i(t_{init}),
 \end{eqnarray}
 where $H_D \equiv \dot{D}(t)/D(t)$. 
  
We integrate these equations numerically to find the axis ratios of an 
ellipsoid at the time of shell crossing in terms of its initial $e$ and $p$.
Fig.~\ref{fig:alphacompare} compares the result of this calculation to 
the Zeldovich approximation. Here we plot the smallest ratio of the 
principal axes $\alpha=\left(J_3/J_1\right)^{1/2}$ at the present time 
calculated from the ellipsoidal evolution of~\citet{1996ApJS..103....1B} 
( $\alpha_{BM}$) against the corresponding ratio calculated 
from the Zeldovich approximation ($\alpha_{Zeld}$), for different values 
of the other ratio of axes $\beta=\left(J_2/J_1\right)^{1/2}$ computed 
using the Zeldovich approximation in different panels. 
The blue dots are for the linear approximation for the evolution of the 
outside tidal field and the magenta dots for the non-linear model for
external tides
(see~\citet{1996ApJS..103....1B} for a detailed 
discussion on these choices). The solid line shows the curve 
$\alpha_{{BM}}=\alpha_{Zeld}$. Since we must have 
$0<\alpha<\beta<1$ the dots only extend to $\alpha<\beta$. 
It shows that the ellipsoidal collapse approximation is very similar to
the Zeldovich approximation for voids. 
\begin{figure*}
\begin{center}
\includegraphics[width=0.9\textwidth]{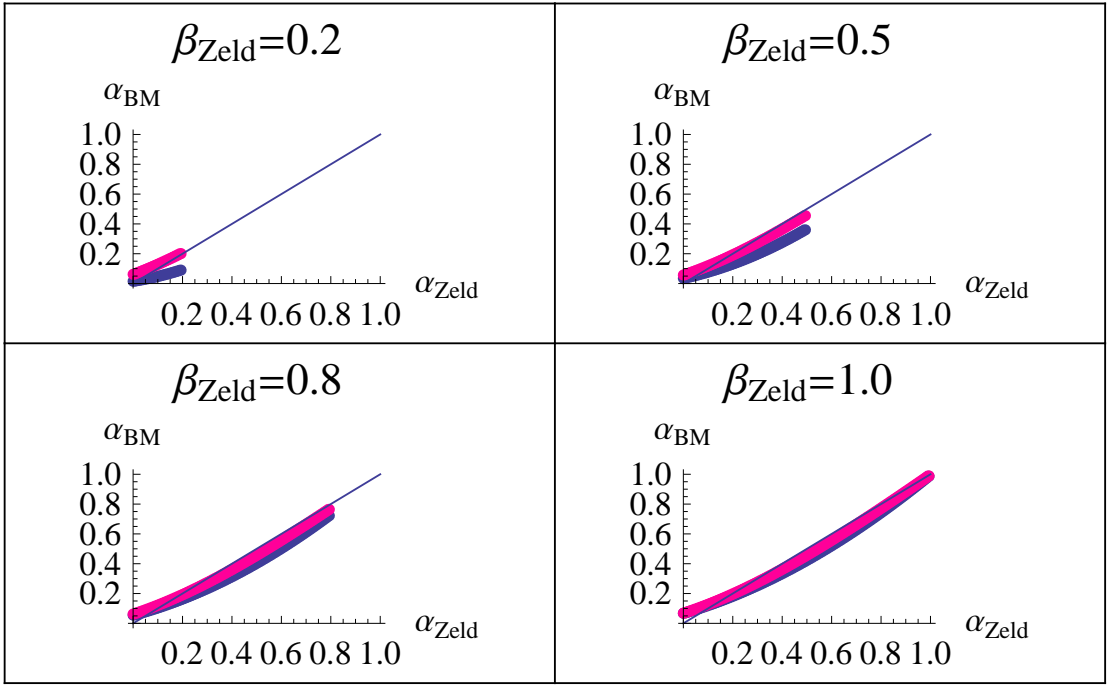}
\end{center}
\caption{
A comparison of evolution of the axes 
ratios $\alpha$ and $\beta$ of the void ellipsoid computed using 
the linear tide approximation of ~\citet{1996ApJS..103....1B} (blue dots),
the non-linear tide approximation (magenta dots) against the ratios
computed using Zeldovich approximation for different values of the 
axes ratio $\beta$} 
\label{fig:alphacompare}
\end{figure*}
\end{document}